\documentclass[12pt,preprint]{aastex}

\usepackage{CJK}
\usepackage{natbib}
\usepackage{graphicx}
\usepackage{longtable, lscape}

\begin{document}
\begin{CJK}{UTF8}{gbsn}

\title{Elemental Compositions of Two Extrasolar Rocky Planetesimals
\thanks{Some of the data presented herein were obtained at the W.M. Keck Observatory, which is operated as a scientific partnership among the California Institute of Technology, the University of California and the National Aeronautics and Space Administration. The Observatory was made possible by the generous financial support of the W.M. Keck Foundation.} }

\author{S. Xu(许\CJKfamily{bsmi}偲\CJKfamily{gbsn}艺)\altaffilmark{a}, M. Jura\altaffilmark{a}, D. Koester\altaffilmark{b}, B. Klein\altaffilmark{a}, B. Zuckerman\altaffilmark{a}}
\altaffiltext{a}{Department of Physics and Astronomy, University of California, Los Angeles CA 90095-1562; sxu@astro.ucla.edu, jura@astro.ucla.edu, kleinb@astro.ucla.edu, ben@astro.ucla.edu}
\altaffiltext{b}{Institut fur Theoretische Physik und Astrophysik, University of Kiel, 24098 Kiel, Germany; koester@astrophysik.uni-kiel.de}

\begin{abstract}
We report Keck/HIRES and HST/COS spectroscopic studies of extrasolar rocky planetesimals accreted onto two hydrogen atmosphere white dwarfs, G29-38 and GD 133. In G29-38, 8 elements are detected, including C, O, Mg, Si, Ca, Ti, Cr and Fe while in GD 133, O, Si, Ca and marginally Mg are seen. These two extrasolar planetesimals show a pattern of refractory enhancement and volatile depletion. For G29-38, the observed composition can be best interpreted as a blend of a chondritic object with some refractory-rich material, a result from post-nebular processing. Water is very depleted in the parent body accreted onto G29-38, based on the derived oxygen abundance. The inferred total mass accretion rate in GD 133 is the lowest of all known dusty white dwarfs, possibly due to non-steady state accretion. We continue to find that a variety of extrasolar planetesimals all resemble to zeroth order the elemental composition of bulk Earth. 
\end{abstract}


\section{Introduction}

Based upon models of planet formation, we understand the variety of elemental compositions of planetesimals as a familiar three step process \citep{McSweenHuss2010}. (i) Under nebular condensation, incorporation of an element into the planetesimal is a function of local temperature and pressure. (ii) Differentiation often occurs, redistributing all elements within the planetesimal; lithophile elements (Al, Ca, Ti) are enhanced in the crust while siderophile elements (Fe, Mn, Cr, Ni) settle into the core. (iii) Collisions lead to stripping and blending of cores and crust, redistributing elements within the entire planetary system. In the solar system, chondrites are a direct consequence of nebular condensation, i.e. step (i); achondrites and primitive achondrites have experienced post-nebular processing, i.e. steps (ii) and (iii) \citep{ONeillPlame2008}. The study of externally-polluted white dwarfs provides invaluable information about the elemental compositions of extrasolar rocky planetesimals, directly testing these models and contrasting with solar system objects \citep{Jura2013a,JuraYoung2014}. 

The current picture is that beyond a few AUs, a large fraction of extrasolar planetesimals can survive to the white dwarf phase \citep{Jura2008}. From dynamical rearrangement during the post-AGB phase, some planetesimals can be perturbed into the tidal radius of the white dwarf and subsequently ``pollute" its pure hydrogen or helium atmosphere \citep{DebesSigurdsson2002, Jura2003, Bonsor2011, Debes2012a,Veras2013}. Calcium, the most easily detected element from optical surveys, has been identified in over 200 white dwarfs \citep{Zuckerman2003, Zuckerman2010, Koester2005b, Dufour2007, Koester2011}. So far, 30 heavily polluted white dwarfs have been found to show excess infrared radiation coming from the debris of these pulverized planetesimals [e.g. \citet{Mullally2007, Farihi2009, XuJura2012}, and references therein]. These stars always show 10 $\mu$m circumstellar silicate emission features when observed spectroscopically \citep{Reach2005b, Reach2009, Jura2009a}. Orbiting gaseous material has been detected in 9 polluted white dwarfs \citep{Gaensicke2006, Gaensicke2007, Gaensicke2008, Gaensicke2011, Gaensicke2012, Melis2010, Melis2012, Farihi2012a, Debes2012b}. With high-resolution spectroscopic observations, 19 heavy elements have been detected in white dwarf atmospheres, including C, O, Na, Mg, Al, Si, P, S, Ca, Sc, Ti, V, Cr, Mn, Fe, Co, Ni, Cu and Sr \citep{Zuckerman2007, Klein2010, Klein2011, Dufour2010, Dufour2012, Farihi2010b, Melis2010, Vennes2010, Vennes2011a, Zuckerman2011,Jura2012, Gaensicke2012, Xu2013a}.

Theoretical calculations show that extrasolar planetesimals with internal water can survive the red giant stage of their parent star \citep{JuraXu2010}. We can constrain the water mass fraction in extrasolar planetesimals by determining the abundance of accreted hydrogen and/or oxygen in polluted white dwarfs \citep{Klein2010}. By analyzing the hydrogen abundance in an ensemble of helium dominated white dwarfs, \citet{JuraXu2012} found that water is less than 1\% of the total accreted material. Recently, \citet{Farihi2013} identified a white dwarf which has accreted a large amount of oxygen, in excess of what can be combined into MgO, SiO$_2$, FeO, CaO and Al$_2$O$_3$; they concluded that the accreted asteroid is at least 26\% water by mass. In addition, if there is enough water in the disk, molecular water emission lines might be detectable in the near-infrared, similar to those around T Tauri stars \citep{CarrNajita2008}.

In this paper, we focus on two pulsating ZZ Ceti hydrogen white dwarfs, G29-38 (WD 2326+049) and GD 133 (WD 1116+026). G29-38 is a fascinating white dwarf and a record holder. It is the first and also the closest white dwarf identified with an infrared excess \citep{ZuckermanBecklin1987} and a 10 $\mu$m circumstellar silicate emission feature \citep{Reach2005b,Reach2009}. It is among the very first few hydrogen white dwarfs that were found to be polluted \citep{Koester1997}, which led to the identification of a white dwarf subclass ``DAZ" \citep{Zuckerman2003}. Very recently, G29-38, together with GD 133 and GD 31 are the first white dwarfs with photospheric detections of molecular hydrogen \citep{Xu2013b}. The atmospheric pollution in GD 133 was first reported in the SPY survey \citep{Koester2005b, KoesterWilken2006}. It also has an orbiting dust disk as well as a 10 $\mu$m silicate feature \citep{Jura2007b,Jura2009a}.

The paper is organized as follows. In Section 2, we report data acquisition and reduction methods. In Section 3, we determine stellar parameters for G29-38 and GD 133. The heavy element abundances are reported in Section 4. In Section 5, we compare the composition of the parent body accreted onto G29-38 and GD 133 with solar system objects. In Section 6, we put our results in perspective and conclusions are given in Section 7. 

\section{Observations}

We performed spectroscopic studies of G29-38 and GD 133 with the High Resolution Echelle Spectrometer (HIRES) on the Keck telescope and the Cosmic Origins Spectrograph (COS) on the Hubble Space Telescope. G29-38 was also observed with the NIRSPEC on the Keck Telescope. The observation logs are presented in Table \ref{Tab: Log} and described in detail below.

\begin{table}[hp]	
\begin{center}
\caption{Observation Logs} 
\begin{tabular}{lllcccccc}
\\
\hline \hline
Star	& UT Date 	& Instrument	&	$\lambda$ range	& Exposure	\\
	&	& & ({\AA})	& (sec) \\
\hline
G29-38 	& 2006 Jun 11  & HIRES/red	&  5690-10160	& 7,200\\
      		& 2008 Aug 7	& HIRES/blue	&  3115-5950	& 2,000\\
		& 2010 Oct 17	& COS & 1145-1445 & 1,999\\
		& 2011 Jan 19	& COS & 1145-1445	& 7,035\\
		& 2011 Aug 15	& NIRSPEC	& 27,500-36,000	& 2,160 \\
GD 133 	& 2008 Feb 13	& HIRES/blue	& 3135-5965	& 2,700\\
 		& 2008 Feb 26	& HIRES/red	& 4600-8995	& 2,400\\
 		& 2011 May 28	& COS		& 1145-1445 	& 13,460\\
\hline
\label{Tab: Log}
\end{tabular}
\end{center}
\end{table}

\subsection{{\it Keck}/HIRES Optical Spectroscopy}

The optical data were acquired with HIRES \citep{Vogt1994} on the Keck I telescope at Mauna Kea Observatory under good weather conditions except for the night of Aug 7, 2008, where high cirrus clouds were present that caused 2-3 magnitudes of extinction. The C5 slit with a width of 1$\farcs$148 was used for all observations. The spectral resolution is $\sim$ 40,000 as measured from the Th-Ar lamps. 

The MAKEE software\footnote{MAKEE Keck Observatory HIRES Data Reduction Software, http://www2.keck.hawaii.edu/inst/common/makeewww/} was used to extract the spectra from the flat-fielded two-dimensional image of each exposure with the trace of a bright calibration star. Wavelength calibration was performed using the standard Th-Ar lamps. Following \citet{Klein2010, Klein2011}, we used IRAF to normalize the spectra and combine echelle orders.  When multiple exposures were present, each exposure was processed separately based on steps outlined above and combined afterwards, weighted by their count rate. For GD 133, there was second order contamination in 8200-9000 {\AA} region and we followed \citet{Klein2010} to calibrate and extract that part of the spectrum. For both stars, the final spectra were continuum-normalized but not flux calibrated. The signal-to-noise ratio (S/N) for G29-38 is 50-90 shortward of 3850 {\AA} and 90-210 for longer wavelengths. For GD 133, the S/N is 30-60 shortward of 3900 {\AA} and longward of 6000 {\AA} and 60-110 for the rest.

\subsection{{\it HST}/COS Ultraviolet Spectroscopy}

G29-38 and GD 133 were observed as part of the HST cycle 18 program 12290, ``Do Rocky Extrasolar Minor Planets Have a Composition Similar to Bulk Earth?". G29-38 was observed at two different times due to the malfunction of a gyro during part of the first observation. Instrument configuration and data reduction procedures were described in \citet{Xu2013b}. Following \citet{Jura2012}, we extracted night-time portions of the data to remove geocoronal O I emission lines near 1304 {\AA}. For GD 133, there were 4711 sec of useful night time data and the S/N of the unsmoothed spectrum is 8 around O I 1304 {\AA}. For G29-38, there are only 400 sec of effective night time exposure and the data were not used for the analysis. 

\subsection{{\it Keck}/NIRSPEC Infrared Spectroscopy}

G29-38 was observed with the NIRSPEC \citep{McLean1998, McLean2000} on the Keck II telescope in low resolution R $\sim$ 2000 spectroscopy mode and a central wavelength of 3 $\mu$m. The slit size was chosen to be 42 $\times$ 0{\farcs}57, matching the average seeing of the night around 0{\farcs}6. Exposures were 60 sec each; 60 co-added images with a 1 sec frame time. The target was observed at two nod positions; a complete set includes an ABBA nod pattern and has a total on target time of 4 minutes. After 3-5 sets of observations on G29-38, an equal number of sets were taken on the calibration star HD 222749 (B9V) to remove telluric features and instrument transmission features. 

\begin{figure}[hp]
\plotone{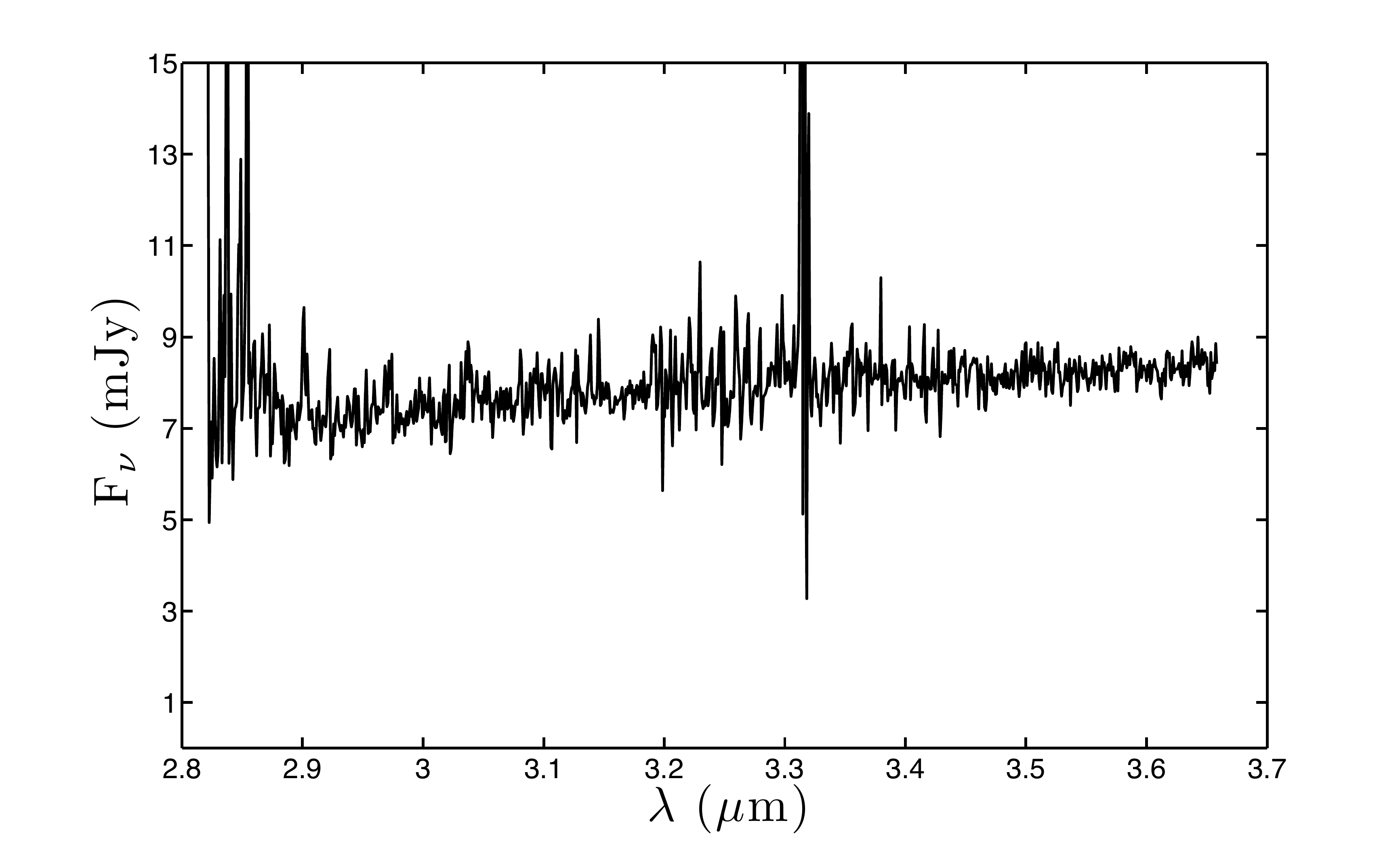}
\caption{Keck/NIRSPEC KL band spectrum for G29-38 and flux calibrated to IRAC 3.6 $\mu$m photometry. The spectral resolution is 2000 and the data are neither binned nor smoothed. The noisy region shortward of 2.9 $\mu$m is due to bright sky background and the emission around 3.3 $\mu$m is from an instrumental artifact. The spectrum is generally featureless with a gentle increase towards longer wavelength, consistent with the dust disk model. }
\label{Fig: NIRSPEC}
\end{figure}

All spectroscopic reductions were made using the REDSPEC software\footnote{NIRSPEC Data Reduction with REDSPEC, http://www2.keck.hawaii.edu/inst/nirspec/redspec.html} following procedures outlined in \citet{McLean2003}, which includes corrections of nonlinearity in the spatial and spectral dimensions, wavelength calibrations with the Ne and Ar lamps, extraction of the spectra and removal of telluric features and the instrument response function. To restore the spectral slope, the spectrum is multiplied by a black body curve of 9150 K, the temperature of the calibration star. The last step is to flux calibrate the spectrum to the IRAC 3.6 $\mu$m flux \citep{Farihi2008a}. The final spectrum is shown in Figure \ref{Fig: NIRSPEC}; it has a higher spectral resolution and S/N than previous near-infrared data from the IRTF \citep{Tokunaga1990} and Gemini \citep{Farihi2008a}.

\section{Model Atmosphere and Stellar Parameters}

Synthetic white dwarf model atmospheres were computed with basic input parameters including effective temperature, T$_{eff}$, surface gravity, g, and atmospheric abundances of heavy elements. The computed model spectra presented here are a new grid with two major changes compared to previous work in \citet{Koester2009a, Koester2010}. (i) The mixing-length parameter ML2/$\alpha$ is taken as 0.8, which is now the preferred value of the Montreal group \citep{Tremblay2010}. (ii) New Stark broadening data are used \citep{TremblayBergeron2009}. The adopted T$_{eff}$ and g are shown in Table \ref{Tab: Parameters} and elemental abundances in Table \ref{Tab: Abundances}. Below we describe the fitting process in detail. Fortunately, precise stellar parameters are not essential for our analysis because we are most interested in the relative abundance ratios, which are fairly insensitive to particular models \citep{Klein2011}.  

\begin{table}[htpb]	
\begin{center}
\caption{\bf Adopted Stellar Parameters} 
\begin{tabular}{llccccccc}
\\
\hline \hline
Star	& M$_*$	& T$_{eff}$	& log g	& D	& log M$_{cvz}$/M$_*$$^a$	\\
	& (M$_{\odot}$)& (K)	& (cm s$^{-2}$)	& (pc)	\\
\hline
G29-38	& 0.85	& 11820 $\pm$ 100	& 8.40 $\pm$ 0.10	& 13.6 $\pm$ 0.8	& -13.9\\
GD 133	& 0.66	& 12600 $\pm$ 200	& 8.10 $\pm$ 0.10	& 36.6 $\pm$ 3.2	& -16.2	 \\
\hline
\label{Tab: Parameters}
\end{tabular}
\end{center}
{\bf Notes.} $^a$ M$_{cvz}$ is the mass of the convection zone. The convection zone of GD 133 is within a Rosseland mean opacity $\sim$ 8, considerably shallower than that of G29-38.
\end{table}

\subsection{G29-38}

G29-38 has a parallax $\pi$= 0.0734 $\pm$ 0.0040 arcsec \citep{vanAltena2001} as well as UBVRI photometry \citep{Holberg2008a}. Its infrared photometry was not used for the fitting due to contamination from the dust disk \citep{ZuckermanBecklin1987}. Additional data were also used for the analysis, including the HST/FOS spectra, two optical spectra from the 2.2m Calar Alto telescope \citep{Koester1997} and two spectra from the VLT/UVES \citep{Koester2005b, Koester2009a}.

\begin{table}[htpb]	
\begin{center}
\caption{Final Atmospheric Abundances} 
\begin{tabular}{lcclccllll}
\\
\hline \hline
	& \multicolumn{3}{c}{G29-38}	& \multicolumn{3}{c}{GD 133} \\
Z	& [Z/H]$^a$	& t$_{set}$	&	$\dot{M}$(Z$_i$)$^b$	& [Z/H]$^a$	& t$_{set}$	&	$\dot{M}$(Z$_i$)$^b$\\
	&	& (10$^{-1}$ yr)	& (g s$^{-1}$) 	& & (10$^{-3}$ yr)	& (g s$^{-1}$)\\
\hline
C 	& -6.90 $\pm$ 0.12   &	7.8	& 1.2 $\times$ 10$^6$		& $<$ -7.9		& 5.3		& $<$ 7.6 $\times$ 10$^4$\\
N	& $<$ -5.7			&	6.4	& $<$ 2.6 $\times$ 10$^7$	& $<$ -5.8		& 3.4		& $<$ 1.7 $\times$ 10$^7$\\
O	& -5.00 $\pm$ 0.12	&	4.5	& 2.2 $\times$ 10$^8$ 		& -6.00 $\pm$ 0.11 & 2.4	& 1.8 $\times$ 10$^7$\\
Na	& $<$ -6.7			&	2.1	& $<$ 1.3	 $\times$ 10$^7$	& $<$ -6.3		& 3.7		& $<$ 8.2 $\times$ 10$^6$\\
Mg	& -5.77 $\pm$ 0.13	&	2.5	& 9.8 $\times$ 10$^7$		& -6.5:	 & 9.2	& 2.2 $\times$ 10$^6$:\\
Al	& $<$ -6.1			&	3.4	& $<$ 3.8 $\times$ 10$^7$ 	& $<$ -5.7		& 6.4		& $<$ 2.3 $\times$ 10$^7$\\
Si	& -5.60 $\pm$ 0.17	&	4.6	& 9.4 $\times$ 10$^7$		& -6.60 $\pm$ 0.13 & 5.5	& 3.4 $\times$ 10$^6$\\
S	& $<$ -7.0			&	4.1	& $<$ 4.8 $\times$ 10$^6$	& $<$ -7.0		& 2.9		& $<$ 3.0 $\times$ 10$^6$\\
Ca	& -6.58 $\pm$ 0.12	&	2.0	& 3.1 $\times$ 10$^7$ 		& -7.21 $\pm$ 0.13 & 6.2	& 1.1 $\times$ 10$^6$\\
Ti	& -7.90 $\pm$ 0.16	&	2.7	& 1.4 $\times$ 10$^6$		& $<$ -8.0		& 5.3		& $<$ 2.4 $\times$ 10$^5$\\
Cr	& -7.51 $\pm$ 0.12	&	2.4	& 4.0 $\times$ 10$^6$		& $<$ -6.8		& 4.3 	& $<$ 5.1 $\times$ 10$^6$ \\
Mn	& $<$ -7.2			&	2.2	& $<$ 9.5 $\times$ 10$^6$	& $<$ -7.0		& 4.2		& $<$ 3.5 $\times$ 10$^6$ \\
Fe	& -5.90 $\pm$ 0.10	&	2.1	& 2.0 $\times$ 10$^8$		& $<$ -5.9		& 3.6		& $<$ 5.2 $\times$ 10$^7$\\
Ni	& $<$ -7.3			&	1.9	& $<$ 9.6 $\times$ 10$^6$	& $<$ -7.0		& 3.1		& $<$ 5.1 $\times$ 10$^6$\\
\\
Total$^c$	&				&		&	6.5 $\times$ 10$^8$ & & & 2.4 $\times$ 10$^7$\\
\hline
\label{Tab: Abundances}
\end{tabular}
\end{center}
{\bf Notes.} See Table \ref{Tab: Lines} for details. \\
$^a$ [X/Y] = log n(X)/n(Y), the logarithmic number ratio of the abundance of element X relative to the abundance of Y.\\
$^b$ The instantaneous mass accretion rate of an element into the white dwarf's atmosphere (see section 5). This is calculated by dividing the mass of an element currently in the convection zone with its settling time.\\
$^c$ The total accretion rate including all elements with positive detections.
\end{table}

\begin{figure}[hp]
\plotone{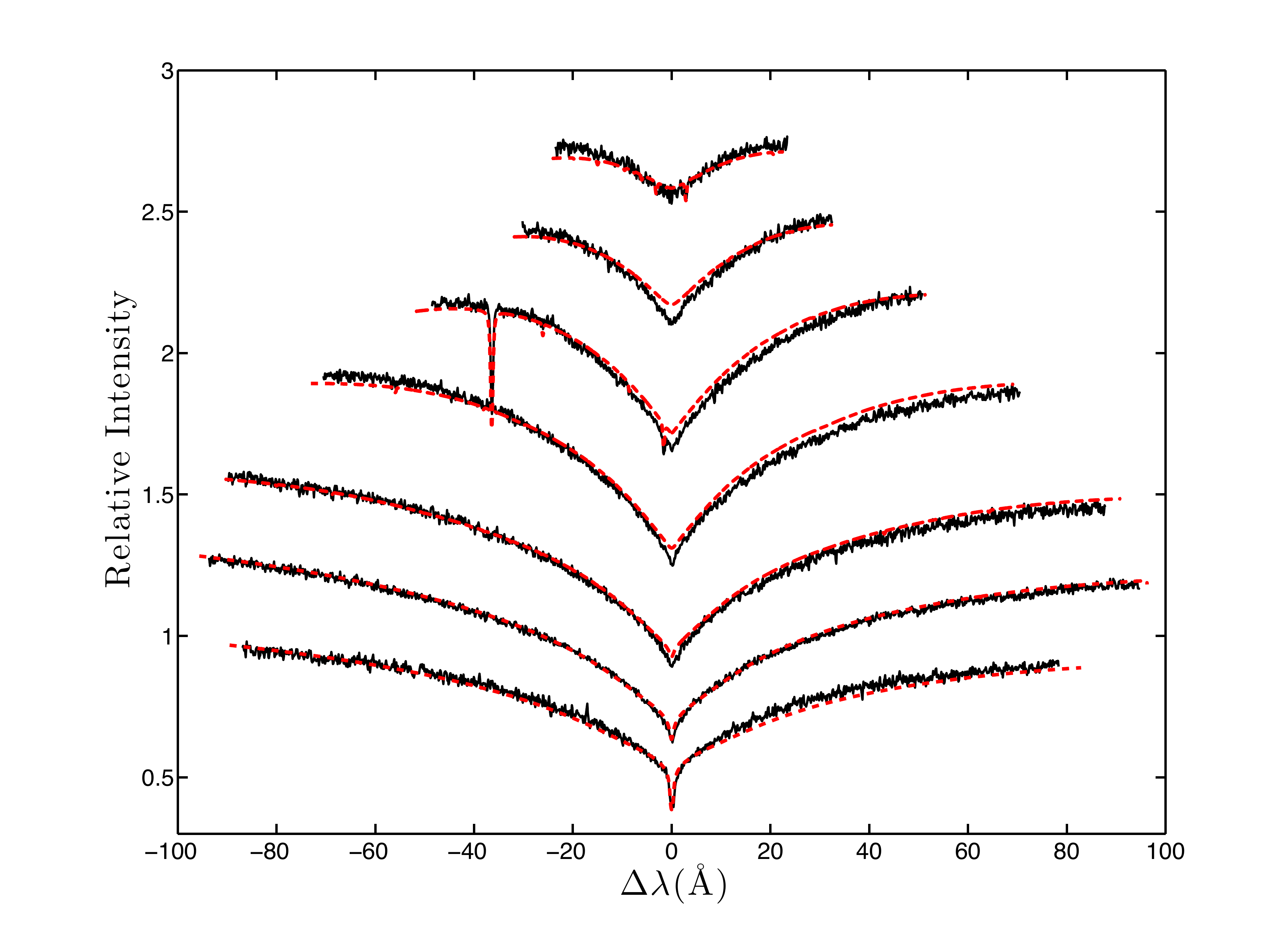}
\caption{Model fits (red dashed lines) to Balmer lines, including H$\alpha$ to H$\eta$ from bottom to top with T$_{eff}$ = 11,820 K, log g=8.40 for G29-38. Each line is offset by 0.3 in relative intensity for clarity. The underlying spectrum in black is from the SPY survey \citep{Koester2005b, Koester2009a}. Ca II K-line is seen at the left wing of H$\epsilon$ with $\Delta$$\lambda$ of -37 {\AA}. }
\label{Fig: BalmerLines}
\end{figure}

The surface gravity of G29-38 can be tightly constrained from the parallax. For any reasonable effective temperature within the instability strip the gravity has to be in the interval 8.30-8.50 with the most consistent solution of 8.40. Varying the parallax within the quoted error of 0.004 arcsec shifts the optimum log g value by 0.05 dex. Holding gravity as a fixed value, we are able to derive T$_{eff}$; with all available observing data, the best solution is listed in Table \ref{Tab: Parameters}. The fits to Balmer lines from H$\alpha$ to H$\eta$ are shown in Figure \ref{Fig: BalmerLines}. The higher order Balmer lines in the model are not as deep as observed.  G29-38 is a pulsating ZZ Ceti white dwarf and the velocity fields tend to cause line profiles to be broader and shallower \citep{KoesterKompa2007}. But this effect is only relevant for the innermost cores within 1 {\AA} and does not influence the parameter determinations. The problem can be solved by adopting a lower surface gravity but this contradicts the parallax measurement. Assuming the parallax is correct, the disagreement could indicate a problem with our implementation of the Balmer line broadening theory \citep{TremblayBergeron2009} and/or the calculation of occupation probabilities based on the prescription of \citet{HummerMihalas1988}\footnote{ The real issue is that G29-38 is in the parameter range where the Balmer line strengths reach their maximum and are rather insensitive to changes in stellar parameters. As a result, we use all available data to derive the stellar parameters. In addition, relative abundance ratios are not strongly dependent on stellar parameters. As illustrated in \citet{Klein2011}, for PG 1225-079, simultaneous changes of 1500 K in temperature and 0.6 dex in log g lead to a mixmum change of 0.1 dex for relative abundances. }. Our newly derived stellar mass is 0.85 M$_{\odot}$, significantly higher than all previous analysis \citep{Koester1997, GIammichele2012} but close to the value of 0.79 M$_{\odot}$ derived from asteroseismology \citep{ChenLi2013}.

\subsection{GD 133}

There is no known parallax for GD 133 and we rely completely on spectroscopic method to derive its stellar parameters. Refitting the SPY spectra \citep{Koester2005b} with our latest model grid gives T$_{eff}$ = 12,729 K and log g = 8.02. GD 133 was also studied by \citet{Gianninas2011}; with a different set of data and model, they derived T$_{eff}$ = 12,600 K and log g = 8.17. The average of the two log g from optical studies is 8.10 and we can derive T$_{eff}$ from fitting the Lyman $\alpha$ profile in the COS spectrum. The formal error of this fitting is extremely small ($\sim$ 10 K) and the quoted error is dominated by the error in log g. The errors listed in Table \ref{Tab: Parameters} for log g and T$_{eff}$ include systematic and statistical errors. The final parameters are close enough to previous values that the new fits are not shown here. 

\section{Atmospheric Abundance Determinations}

To reflect instrumental broadening, the computed model spectra are convolved with the Line Spread Function of COS \citep{Kriss2011} or a Gaussian profile for the HIRES data. Then the abundances of individual elements are determined by comparing the equivalent width (EW) of each spectral line with that from the model spectra \citep{Klein2010, Klein2011}. Compared to helium atmosphere white dwarfs with the same amount of pollution [e.g. \citet{Dufour2012}], the analysis in hydrogen atmosphere white dwarfs is less affected by blending of different absorption lines due to the high continuum opacity of hydrogen atoms. However, molecular hydrogen lines are pervasive in the COS data for both stars. The analysis for GD 133 is less affected because the number density of molecular hydrogen in GD 133 is about 0.4 dex smaller than that in G29-38 \citep{Xu2013b}. For both stars, we present model spectra with and without the contribution from molecular hydrogen. 

Following \citet{Xu2013a}, upper limits to the abundances of elements were estimated by varying the input abundance of an element and comparing the model spectra with data. The presence of numerous molecular hydrogen lines in the COS data complicates this process and the model is not ideal for computing the line strength of individual H$_2$ lines due to the lack of accurate broadening parameters \citep{Xu2013b}. To be conservative, all upper limits obtained from the ultraviolet data were determined by using the model spectra without contributions from H$_2$; the numbers can be lower if molecular hydrogen contributes to a significant portion of the total EW.

All detailed measurements are presented in Table \ref{Tab: Lines} and the final abundances in Table \ref{Tab: Abundances}. For G29-38, there are 27 optical spectral lines identified from 5 different elements and 8 ions, including Mg I, Mg II, Ca I, Ca II, Cr II, Fe I, Fe II and Ti II. The average velocity of all absorption lines, including Doppler shift and gravitational redshift, is 36 $\pm$ 2 km s$^{-1}$. The COS data reveal photospheric detection of C I, O I and Si II with an average velocity of 40 $\pm$ 4 km s$^{-1}$, which agrees with the optical value. We have also detected an interstellar line of Si II at 1260.4 {\AA} and C II at 1334.5 {\AA} with an average velocity of 11 km s$^{-1}$. This is very close to the radial velocity of 9.5 km s$^{-1}$ measured in the Hyades cloud\footnote{According to the dynamical model of the local interstellar medium: http://lism.wesleyan.edu/LISMdynamics.html} \citep{RedfieldLinsky2008}, which lies within 15 pc of the Sun. For GD 133, there are 5 optical spectral lines identified from Ca II and the marginal detection of Mg II as well as 4 ultraviolet lines from Si II and O I. The average velocity shift is 49 $\pm$ 2 km s$^{-1}$ and 58 $\pm$ 4 km s$^{-1}$ for the optical and ultraviolet data, respectively. The difference between these two velocities are most likely due to the absolute wavelength uncertainty of 15 km s$^{-1}$ for the medium resolution grating G130M on {\it COS} (COS Instrument Handbook). Interstellar lines at Si II 1260.4 {\AA}, O I 1302.2 {\AA} and C II 1334.5 {\AA} are also detected with an average velocity of 15 km s$^{-1}$, close to the radial velocities of several nearby clouds, including the Gem Cloud and NGP Cloud. \citep{RedfieldLinsky2008}. In both stars, upper limits were derived for a few elements.

\begin{center}
\begin{longtable}{lllllccccc}
\caption{Measured Equivalent Widths of Photospheric Lines and Abundance Determinations} \\
\hline \hline
	&	&	&	& \multicolumn{2}{c}{G29-38}	& \multicolumn{2}{c}{GD 133} \\
Ion	& $\lambda$$^a$ & E$_{low}$	& log $gf$	& EW	& [Z/H] & EW & [Z/H]\\
	& ({\AA})	& (eV)	&	& (m{\AA}) & & (m{\AA})\\
\hline
\endfirsthead

\multicolumn{6}{c}{Table \ref{Tab: Lines} --- \emph{Continued}} \\ 
\hline
\hline
	&	&	&	& \multicolumn{2}{c}{G29-38}	& \multicolumn{2}{c}{GD 133} \\
Ion	& $\lambda$$^a$ & E$_{low}$	& log $gf$	& EW	& [Z/H] & EW & [Z/H]\\
	& ({\AA})	& (eV)	&	& (m{\AA}) & & (m{\AA})\\
\hline
\endhead

\endfoot
  
 \hline
\endlastfoot
C I	& 1277.6	&	0.01	& -0.40	& 52 $\pm$ 9$^{b-H_2}$	& -6.96 $\pm$ 0.08 & ... & ... \\
C I	& 1329.6	&	0.01	& -0.62	& 42 $\pm$ 8$^{b-H_2}$	& -6.85 $\pm$ 0.08 & ... & ...\\
C II	& 1335.7	&	0.01	& -0.34	& ...	& ...	& $<$ 20$^{b-H_2}$	& $<$ -7.9 \\
C	&	&	&	&	& -6.90 $\pm$ 0.12 & & $<$ -7.9 \\
\\
N I		& 1411.9	& 	3.58	& -1.3	& $<$ 27	& $<$ -5.7	& $<$ 25	& $<$ -5.8 \\
\\
O I	& 1152.2	&	1.97	& -0.27	& 81 $\pm$ 22$^{b-H_2}$	& -5.00 $\pm$ 0.12 & ... & ...\\
O I	& 1304.9	&	0.02	& -0.84	& ...	& ...	& 63 $\pm$ 16	& -6.00 $\pm$ 0.11 \\
O	&	&	&	&	& -5.00 $\pm$ 0.12	&	& -6.00 $\pm$ 0.11 \\
\\
Na I	& 5891.6	&	0	& 0.12	& $<$ 24	& $<$ -6.7 & $<$ 30	& $<$ -6.3\\
\\
Mg I	& 3830.4	& 	2.71	& -0.23	& 14 $\pm$ 3	& -5.62 $\pm$ 0.09 & ... & ...\\
Mg I	& 3833	&	2.71	& 0.12, -0.36	& 21	$\pm$ 5	& -5.89 $\pm$ 0.10 & ... & ...\\
Mg I	& 3839.4	&	2.72	& 0.39	& 35 $\pm$ 4	& -5.86 $\pm$ 0.05 & ... & ...\\
Mg I	& 5174.1	&	2.71	& -0.39	& 31 $\pm$ 4	& -5.65 $\pm$ 0.06 & ... & ...\\
Mg I	& 5185.0	&	2.71	& -0.17	& 42 $\pm$ 4	& -5.72 $\pm$ 0.04 & ... & ... \\
Mg II	& 4482	& 	8.86	& 0.74, 0.59	&	 41 $\pm$ 7	& -5.91 $\pm$ 0.08 & 14 $\pm$ 4	& -6.5: \\
Mg	&	&	&	&	& -5.77 $\pm$ 0.13 & ... & -6.5: \\
\\
Al I	& 3945.1	&	0	& -0.62	& $<$ 16	& $<$ -6.1 	& $<$ 19	& $<$ -5.7\\
\\
Si II	& 1260.4 	&	0	& 0.387	& ...	& ...	& 93 $\pm$ 26	& -6.67 $\pm$ 0.11\\
Si II	& 1264.7	&	0.04	& 0.64	& 320 $\pm$ 64$^{b-Si II}$ & -5.76 $\pm$ 0.09 & 115 $\pm$ 20	& -6.70 $\pm$ 0.07\\
Si II	& 1265.0	&	0.04	& -0.33	& 320 $\pm$ 64$^{b-Si II}$	& -5.76 $\pm$ 0.09 & 51 $\pm$ 12 & -6.60 $\pm$ 0.13\\
Si II	& 1309.3	&	0.04	& -0.43	& 106 $\pm$ 11$^{b-H_2}$	& -5.48 $\pm$ 0.05 &61 $\pm$ 12$^{b-H_2}$	& -6.42 $\pm$ 0.05\\
Si	&	&	&	&	& -5.60 $\pm$ 0.17 & & -6.60 $\pm$ 0.13\\
\\
S I	& 1425.0	&	0	& -0.12	& $<$ 55$^{b-H_2}$	& $<$ -7.0 & $<$ 38$^{b-H_2}$	& $<$ -7.0\\
\\
Ca I	& 4227.9	&	0	& 0.27	& 17 $\pm$ 3	&	-6.67 $\pm$ 0.07 & ... &... \\
Ca II	& 3159.8	&	3.12	& 0.24	& 67 $\pm$ 5	&	-6.69 $\pm$ 0.03 & ... &... \\
Ca II	& 3180.3	&	3.15	& 0.50	& 109 $\pm$ 5	&	-6.61 $\pm$ 0.02 & 37 $\pm$ 4 & -7.23 $\pm$ 0.05 \\
Ca II	& 3182.2	& 	3.15	& -0.46	& 40 $\pm$ 4	&	-6.46 $\pm$ 0.04  & ... &... \\
Ca II	& 3707.1	&	3.12	& -0.48	& 24 $\pm$ 4	&	-6.65 $\pm$ 0.07 & ... &... \\
Ca II	& 3934.8	&	0	& 0.11	& 294 $\pm$ 6 &	-6.65 $\pm$ 0.01 & 154 $\pm$ 13 & -7.08 $\pm$ 0.04 \\
Ca II	& 3969.6	&	0	& -0.2	& 78	$\pm$ 7	&	-6.58 $\pm$ 0.04 & 27 $\pm$ 6 & -7.07 $\pm$ 0.09 \\
Ca II	& 8500.4	&	1.69	& -1.4	& 84 $\pm$ 6	&	-6.34 $\pm$ 0.03 & ... & ... \\
Ca II	& 8544.4	& 	1.69	& -0.46	& 185 $\pm$ 16 &	-6.54 $\pm$ 0.04 & 	42 $\pm$ 7 & -7.33 $\pm$ 0.07 \\
Ca II	& 8664.5	&	1.69	& -0.72	& No Data	& No Data & 28 $\pm$ 4 & -7.34 $\pm$ 0.07 \\
Ca	&	&	&	&	& -6.58 $\pm$ 0.12 & & -7.21 $\pm$ 0.13\\
\\
Ti II	& 3235.4	& 	0.05	& 0.43	& 13 $\pm$ 2	& -8.06 $\pm$ 0.06 & ... & ...\\
Ti II	& 3237.5	& 	0.03	& 0.24	& 16 $\pm$ 6	& -7.88 $\pm$ 0.15 & ... & ...\\
Ti II	& 3350	&	0.61,0.05	& 0.43,0.53	& 31 $\pm$ 4	& -7.95 $\pm$ 0.05 & ... & ...\\
Ti II	& 3362.2	&	0.03	& 0.43	& 11	$\pm$ 2	& -8.12 $\pm$ 0.08 & $<$ 12 & $<$ -8.0\\
Ti II	& 3373.8	&	0.01	& 0.28	& 19 $\pm$ 3	& -7.79 $\pm$ 0.05 & ... & ...\\
Ti II	& 3384.7	& 	0	& 0.16	& 19 $\pm$ 4	& -7.70 $\pm$ 0.09 & ... & ...\\
Ti II	& 3686.3	&	0.61	& 0.13	& 11	$\pm$ 2	& -7.87 $\pm$ 0.08 & ... & ...\\
Ti 	&	&	&	&	& -7.90 $\pm$ 0.16 & ... & $<$ -8.0 \\
\\
Cr II	& 3125.9	& 	2.46	& 0.30	& 13 $\pm$ 4	&	-7.42 $\pm$ 0.11	& No Data	& No Data\\
Cr II	& 3133.0	&	2.48	& 0.42	& 10	$\pm$ 4	&	-7.61 $\pm$ 0.17	& No Data	& No Data\\
Cr II	& 3369.0	&	2.48	& -0.09	&	...	& ...	& $<$ 14	& $<$ -6.8 \\
Cr	&	&	&	&	& -7.51 $\pm$ 0.12	&	& $<$ -6.8\\
\\
Mn II	& 3443.0	& 	1.78	& -0.36	& $<$ 14	& $<$ -7.2 & $<$ 15	& $<$ -7.0\\	
\\
Fe I	& 3582.2	& 	0.86	& 0.41	&	12 $\pm$ 2	& -5.97 $\pm$ 0.06 & ... & ... \\
Fe I	& 3735.9	&	0.86	& 0.32	&	13 $\pm$ 3	& -5.82 $\pm$ 0.09 & ... & ...\\
Fe II	& 1361.4	&	1.67	& -0.52	&	...	&	...	& $<$ 17	& $<$ -5.9  \\
Fe II	& 3228.7	&	1.67	& -1.18	&	12 $\pm$ 3	& -5.79 $\pm$ 0.09 & ... & ...\\
Fe	&	&	&	&	& -5.90 $\pm$ 0.10 &  & $<$ -5.9 \\
\\
Ni II	& 1335.2	&	0.19	& -0.19	&	$<$ 20$^{b-H_2}$	&$<$  -7.3	&  $<$ 24$^{b-H_2}$	& $<$ -7.0\\
\label{Tab: Lines}
\end{longtable}
\end{center}
{\bf Notes.} The adopted abundances are also presented in Table \ref{Tab: Abundances}. \\
$^a$ Wavelengths are in vacuum. Atomic data for absorption lines are taken from the Vienna Atomic Line Database \citep{Kupka1999}. \\
$^b$ Blended line. The contributing element is noted in the superscript and the EW is the sum of the blend.

\subsection{Carbon}

\begin{figure}[hp]
\plotone{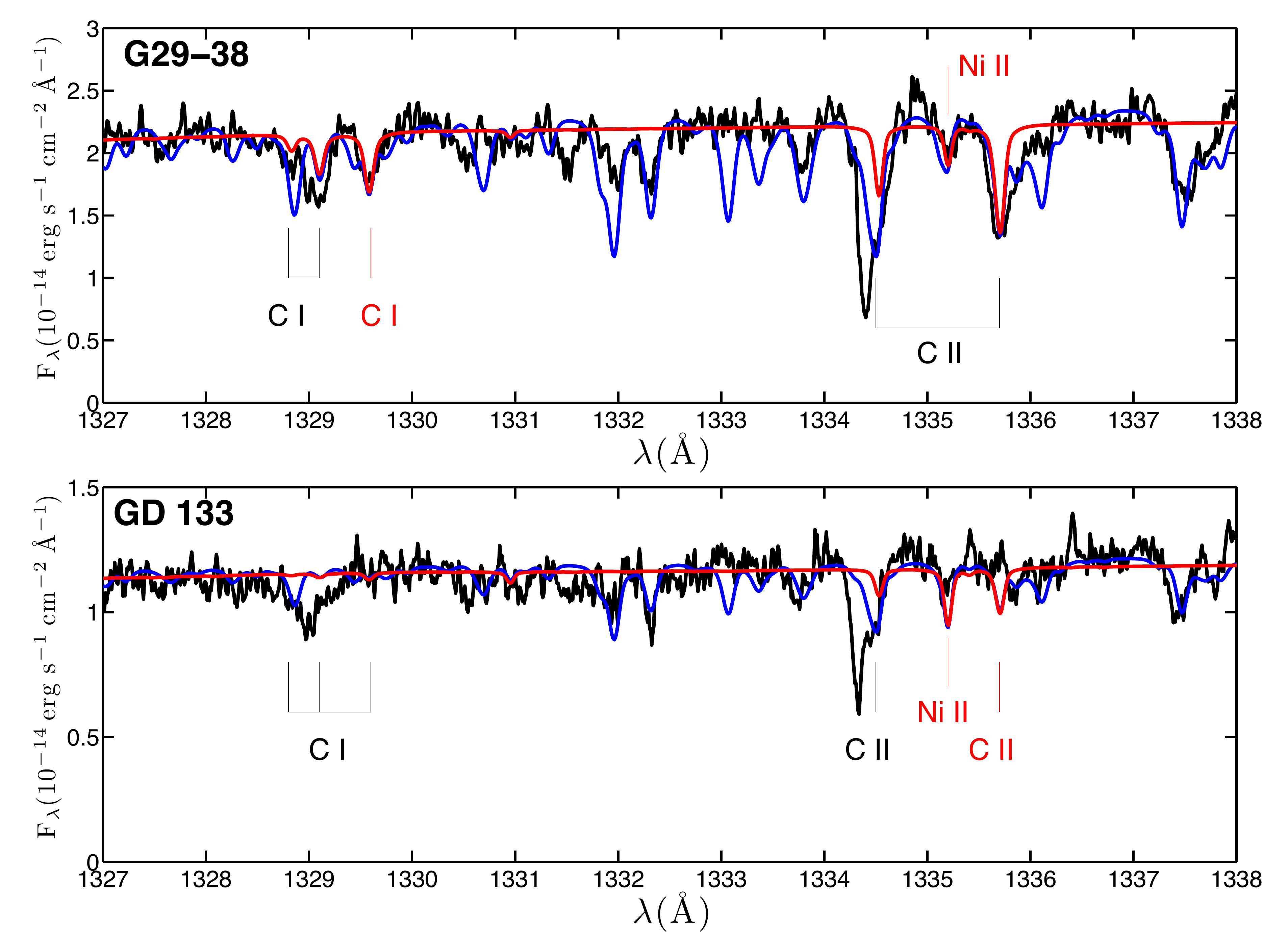}
\caption{Carbon and nickel spectral line region for G29-38 and GD 133. The black line represents HST/COS data smoothed with a 3 pixel boxcar. The red and blue lines represent the model spectra without and with contributions from molecular hydrogen respectively, and with abundances from Table \ref{Tab: Abundances}. Red labels represent lines that are used for abundance determinations in Table \ref{Tab: Lines}. Wavelengths are given in vacuum and the heliocentric reference frame.}
\label{Fig: C_1}
\end{figure}

The carbon lines in the observed wavelength interval can be contaminated by interstellar absorption because they arise from low lying energy levels. In a stellar environment, the C II 1335.7 line is always stronger than the C II 1334.5 line due to its larger statistical weight. However, as shown in Figure \ref{Fig: C_1}, the observed C II 1335.7 line is relatively weaker in both stars; we conclude that these C II lines are mostly interstellar. In addition, the measured velocity for the C II 1334.5 {\AA} line is offset from all other photospheric lines in G29-38 and GD 133. This line also suffers from additional contamination from H$_2$.

\begin{figure}[hp]
\plotone{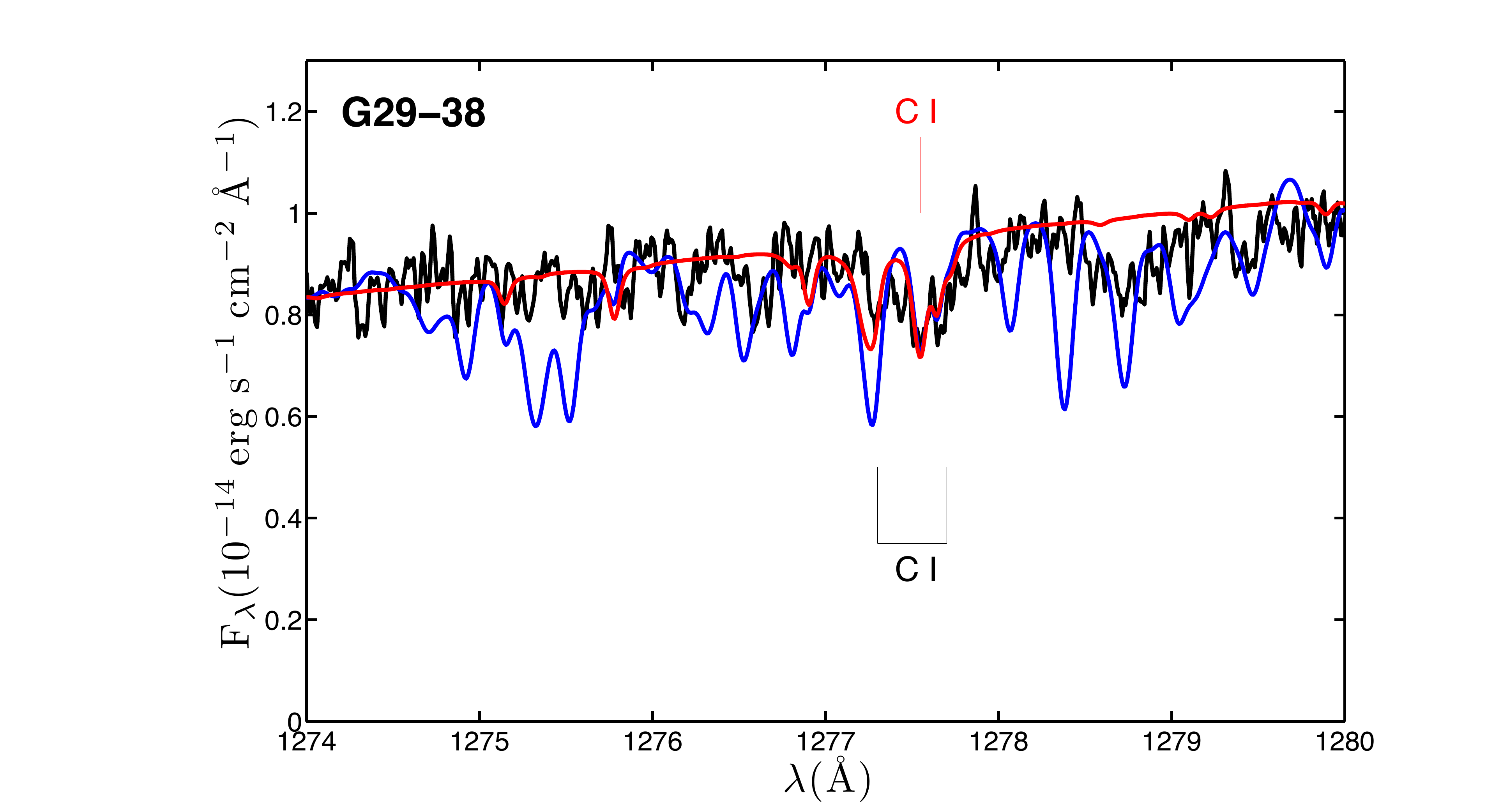}
\caption{Similar to Figure \ref{Fig: C_1} except for C I lines in G29-38. }
\label{Fig: C_G29-38}
\end{figure}

There are C I lines at 1277.6 {\AA} and 1329.6 {\AA} detected in G29-38, as shown in Figures \ref{Fig: C_1} and \ref{Fig: C_G29-38}. The detection is nominally at least 5 $\sigma$ and, in the absence of H$_2$, the derived abundance [C/H] would be -6.90 $\pm$ 0.08. However, due to their proximity to H$_2$ lines, the EW of the C I lines are less certain and we assign a conservative final abundance of -6.90 $\pm$ 0.12. For GD 133, the C II line at 1335.7 {\AA} was used to place the upper limit, which is consistent with the upper bound from the C I 1329.6 {\AA} line in Figure \ref{Fig: C_1}. 

\clearpage

\subsection{Oxygen}

\begin{figure}[hp]
\plotone{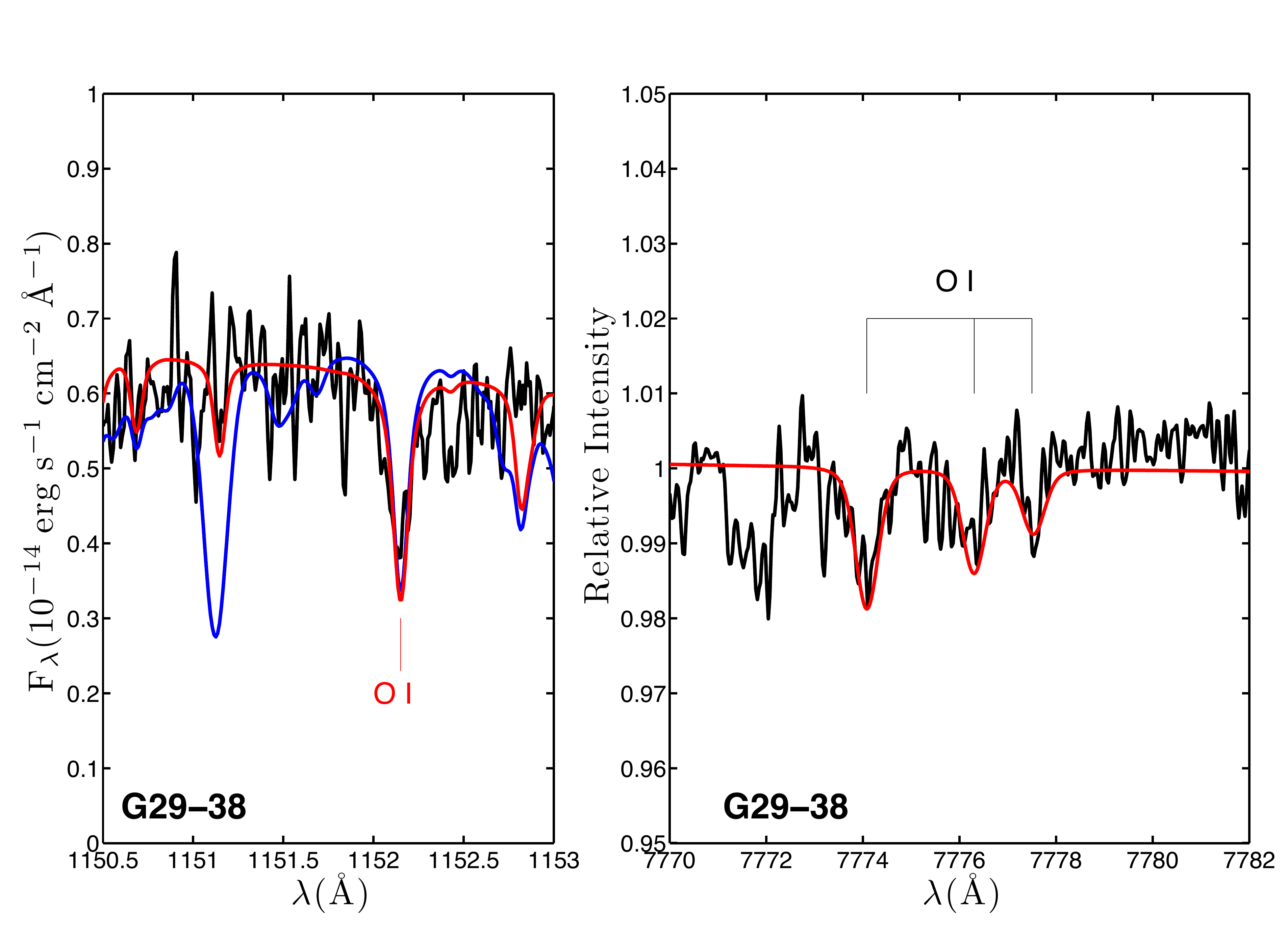}
\caption{Similar to Figure \ref{Fig: C_1} except for the detection of O I in G29-38. The right panel shows Keck/HIRES data smoothed with a 3 pixel boxcar and the spectrum is not flux calibrated. Only one model is plotted for the optical data, which are free from H$_2$ contamination.}
\label{Fig: O_G29-38}
\end{figure}

As presented in Figure \ref{Fig: O_G29-38}, we have detected O I 1152.2 {\AA} in G29-38, which can be reproduced by a model with an oxygen abundance of -5.00. The upper bound from O I triplet lines around 7775 {\AA} also agrees with this result. 

\begin{figure}[htpb]
\plotone{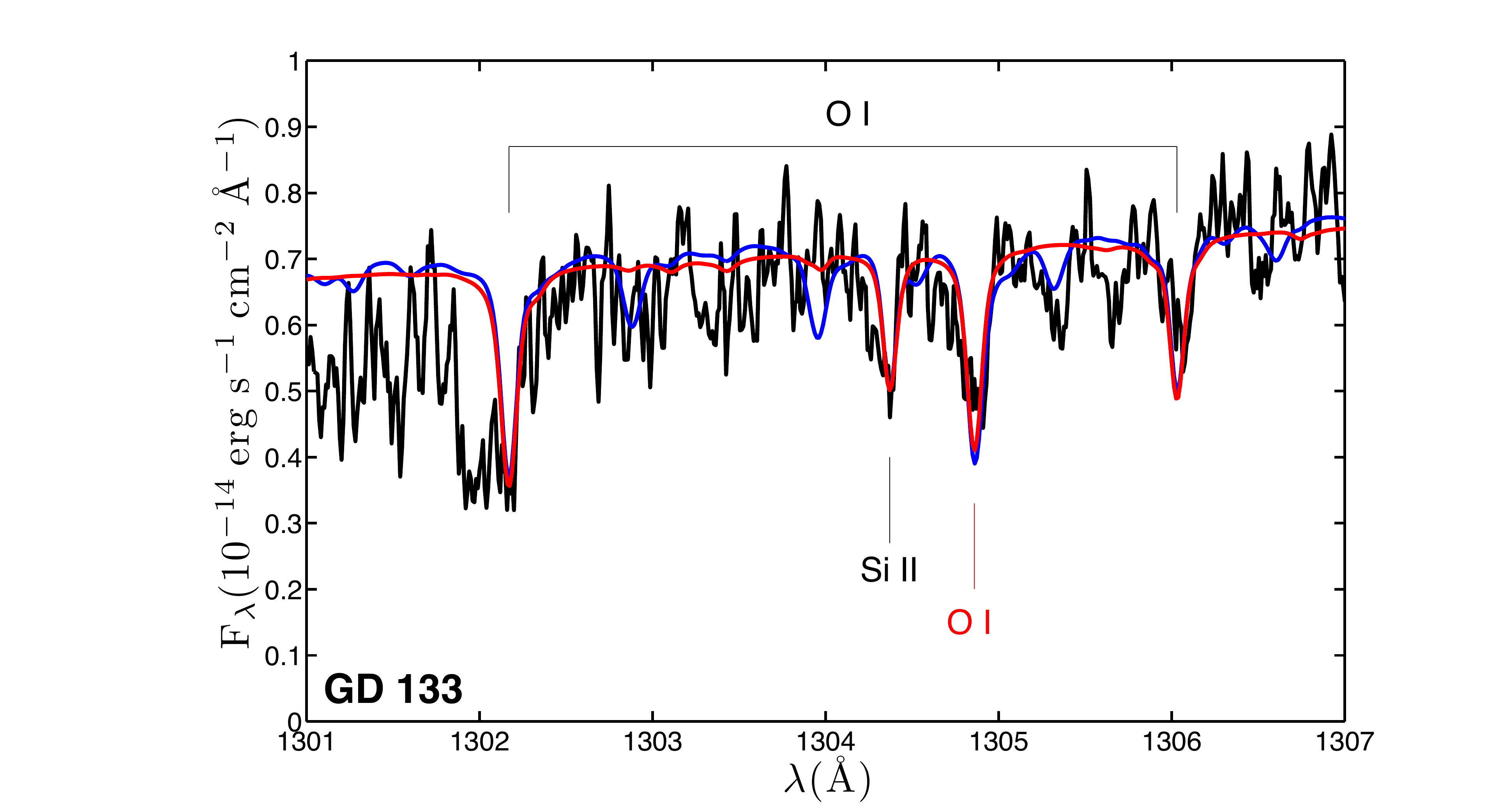}
\caption{Similar to Figure \ref{Fig: C_1} except for the night time portions of COS data of GD 133. The spectrum was smoothed with a 3 pixel boxcar. }
\label{Fig: O_GD133}
\end{figure}

In the wavelength coverage of COS, there are several O I lines around 1300 {\AA}, which can be contaminated from geocoronal emissions. When extracting the night time portions of the data for GD 133, we detected O I lines at 1302.1 {\AA} and 1304.9 {\AA}, as shown in Figure \ref{Fig: O_GD133}. The O I 1302.1 {\AA} line comes from the ground state and there were at least two components, which agree with the interstellar and photospheric contributions, respectively. We use the O I 1304.9 {\AA} line and a model with [O/H] = -6.00 can successfully reproduce the data. This derived abundance is consistent with the strength of the O I 1302.1 {\AA} and O I 1306.0 {\AA} lines in Figure \ref{Fig: O_GD133}. 

\subsection{Magnesium}

\begin{figure}[hp]
\plotone{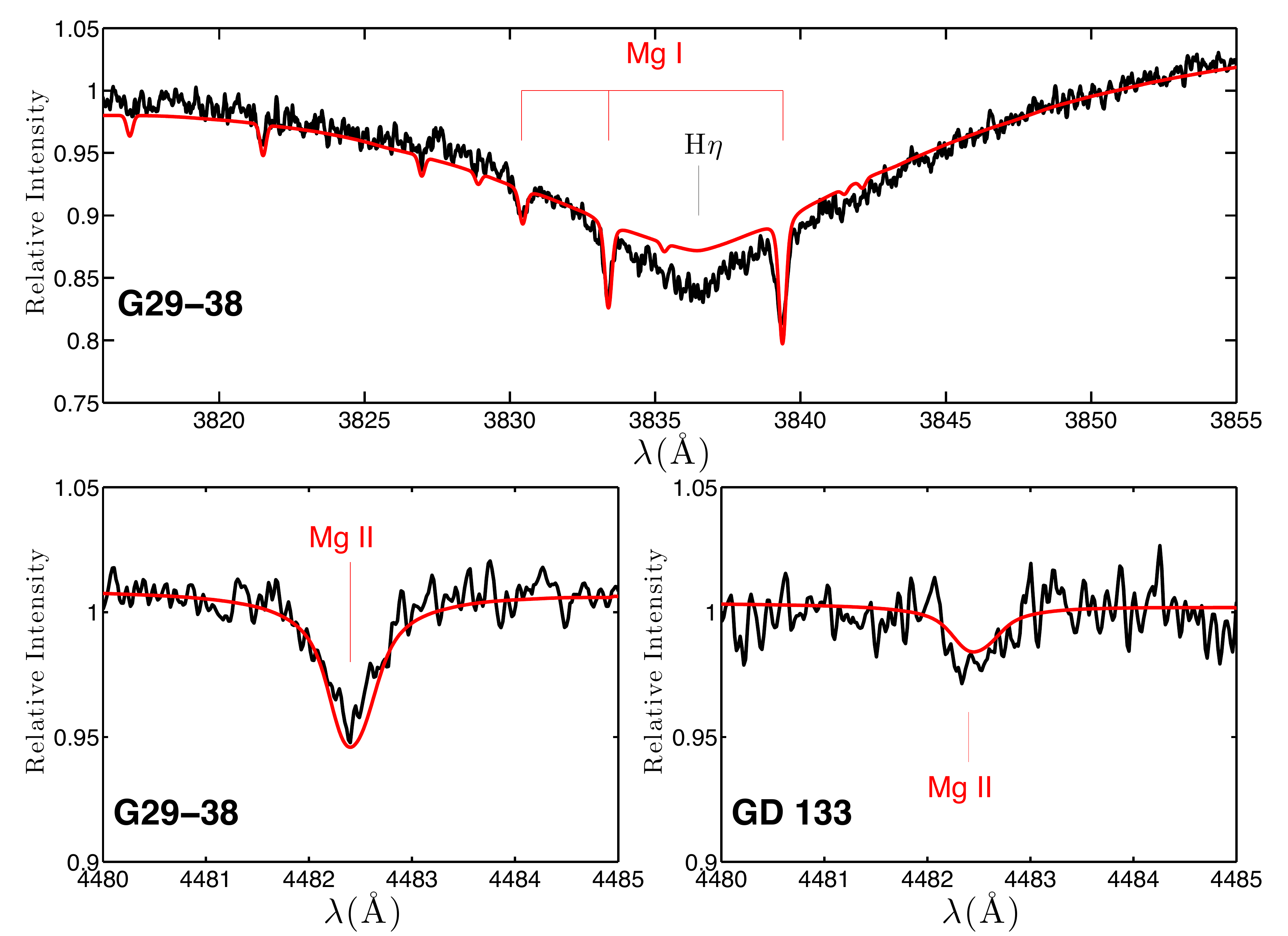}
\caption{Similar to Figure \ref{Fig: C_1} except for magnesium lines in the Keck spectra for G29-38 and GD 133. The spectrum in the top panel was smoothed with a 7 pixel boxcar while the ones in the lower panels are averaged by a 3 pixel boxcar. The fit to H$\eta$ in G29-38 is not ideal (see discussion in section 3.1) but we were still able to directly compare the EWs of Mg I lines in the data with the model. Detection of Mg in GD 133 is marginal.}
\label{Fig: Mg}
\end{figure}

Magnesium is detected only in the optical data, as shown in Figure \ref{Fig: Mg}. In G29-38, the magnesium abundance was derived from a total of 7 spectral lines from both Mg I and Mg II. In GD 133, only the Mg II 4482 {\AA} doublet is marginally seen and we were able to derive a tentative magnesium abundance.

\subsection{Silicon}

\begin{figure}[hp]
\plotone{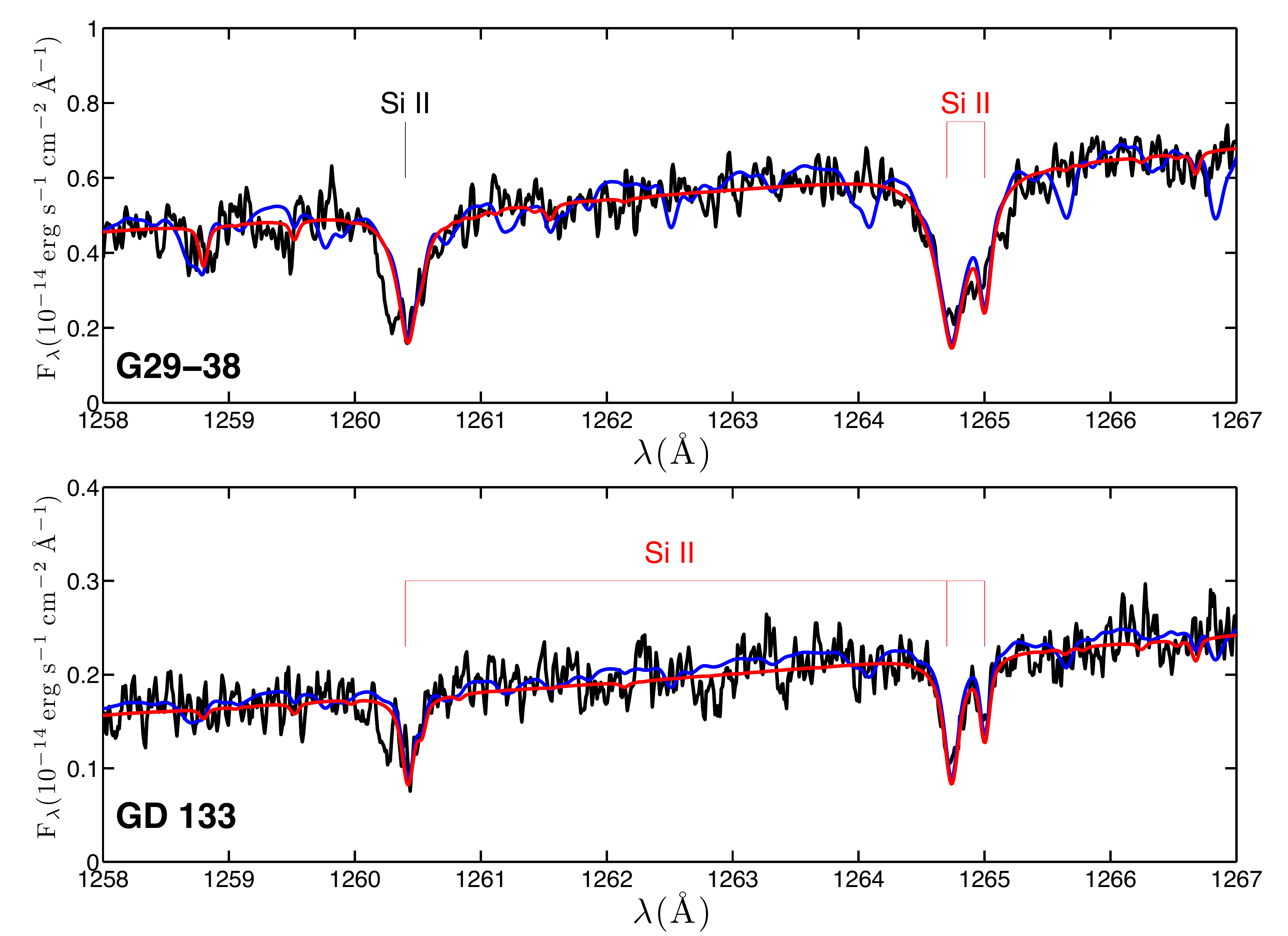}
\caption{Similar to Figure \ref{Fig: C_1} except for silicon lines. The Si II line at 1260.4 {\AA} comes from the ground state and can be only used for the analysis of GD 133 because it is well separated from the interstellar absorption line.}
\label{Fig: Si_1}
\end{figure}

Several ultraviolet silicon lines are used for the analysis, as shown in Figure \ref{Fig: Si_1}. Si II 1260.4 {\AA} was not used for G29-38 because it is blended with an interstellar line. For G29-38, because Si II 1264.7 {\AA} and 1265.0 {\AA} lines are blended with each other, we report the total EW of this feature in Table \ref{Tab: Lines}. Si II 1309.3 {\AA} can be blended with adjacent H$_2$ lines but the silicon line was readily detected and used in the analysis.  For GD 133, the photospheric lines are weaker and easier to deblend. We were able to measure the EW of Si II lines at 1260.4 {\AA}, 1264.7 {\AA} and 1265.0 {\AA} individually to derive the final abundance.

The Si II line at 1260.4 {\AA} in GD 133 is the only readily detected interstellar line that is well separated with the photospheric line. And the EW is 48 $\pm$ 21 m{\AA}, corresponding to a Si II column density of 3.8 $\times$ 10$^{12}$ cm$^{-2}$ in that direction, which is comparable to typical ISM values \citep{Lehner2003}.

\clearpage

\subsection{Calcium}

\begin{figure}[hp]
\plotone{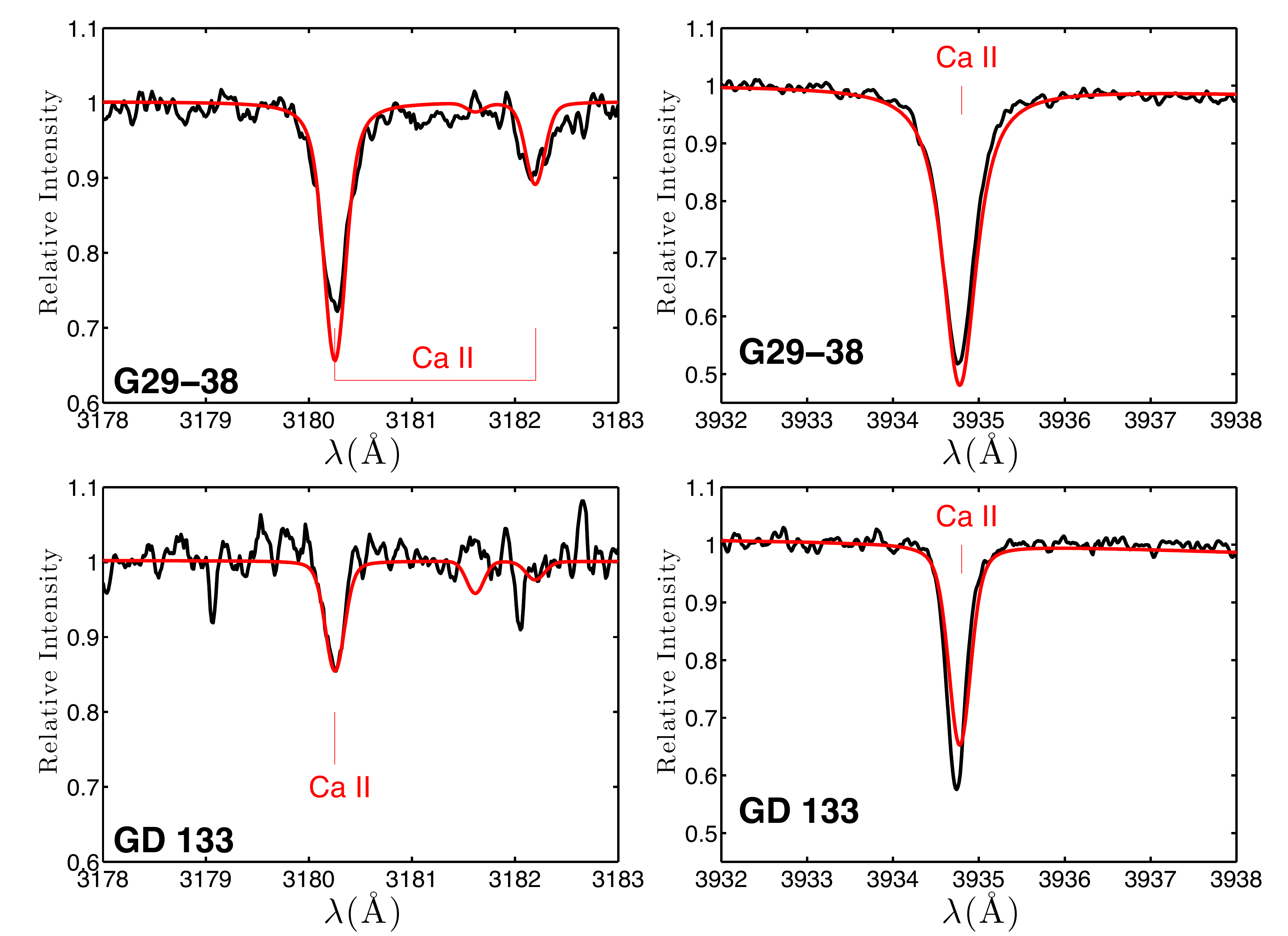}
\caption{Similar to Figure \ref{Fig: C_1} except for calcium lines in the Keck/HIRES data. All spectra were smoothed by a five-point boxcar average.}
\label{Fig: Ca}
\end{figure}

\citet{VonHippelThompson2007} found the EW of the Ca II K-line in G29-38 increased by 70\% in 3 months, which they attributed to a variable mass accretion rate. However, a follow-up study of monitoring the Ca II H \& K, Ca I 4227.9 {\AA} and Mg II 4482 {\AA} lines found no significant variability \citep{DebesMorales2008}. Subsequently, \citet{Thompson2010} found that the EW of the Ca II K-line might be variable by a few percent and concluded that the pulsation model favors polar rather than equatorial accretion. In the present study, we are not concerned about EW uncertainty on a few percent level. There are quite a few calcium lines present in the HIRES spectra in both stars, as listed in Table \ref{Tab: Lines} and shown in Figure \ref{Fig: Ca}. Our measured EWs of the Ca II H \& K and Ca I 4227.9 lines agree within the uncertainties reported in \citet{DebesMorales2008} and \citet{Thompson2010}. In GD 133, there are five Ca II lines detected and [Ca/H] = -7.21. We measured an EW of 154 $\pm$ 13 m{\AA} for the Ca II K-line, which lies within the uncertainty of the EW of 135 m{\AA} from VLT/UVES data but the derived abundance is different due to the updated model atmosphere calculations \citep{Koester2005b, KoesterWilken2006}. 

\subsection{Titanium}

\begin{figure}[hp]
\plotone{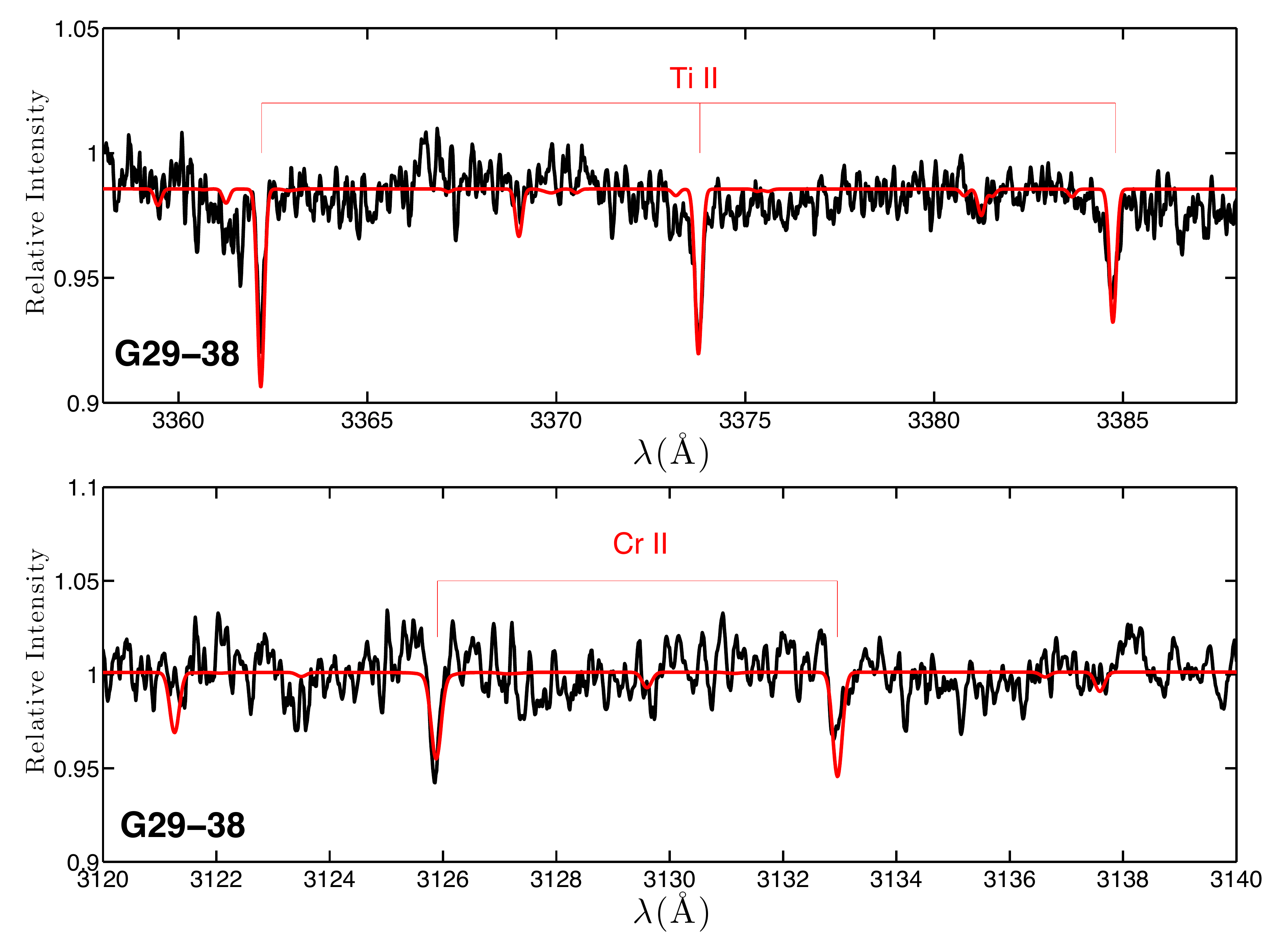}
\caption{Similar to Figure \ref{Fig: C_1} except for titanium and chromium lines. Both spectra were smoothed by a seven-point boxcar average.}
\label{Fig: TiCr}
\end{figure}

Seven titanium lines were detected in the Keck/HIRES spectrum of G29-38 and three are displayed in Figure \ref{Fig: TiCr}. Although most titanium lines have EWs smaller than 20 m{\AA}, they are readily detected due to the high S/N of the data. In GD 133, the Ti II 3362.2 {\AA} line was used to derive the upper limit.

\subsection{Chromium}

In G29-38, there are two chromium lines at 3125.9 {\AA} and 3133.0 {\AA} in the observed wavelength interval as shown in Figure \ref{Fig: TiCr}. The detection of each individual line is marginal but the presence of lines at two correct wavelengths makes the identification more convincing. In GD 133, these two chromium lines fall off the edge of echelle orders and the next strongest chromium line at 3369.0 {\AA} was used to derive the upper limit.

\subsection{Iron}

\begin{figure}[hp]
\plotone{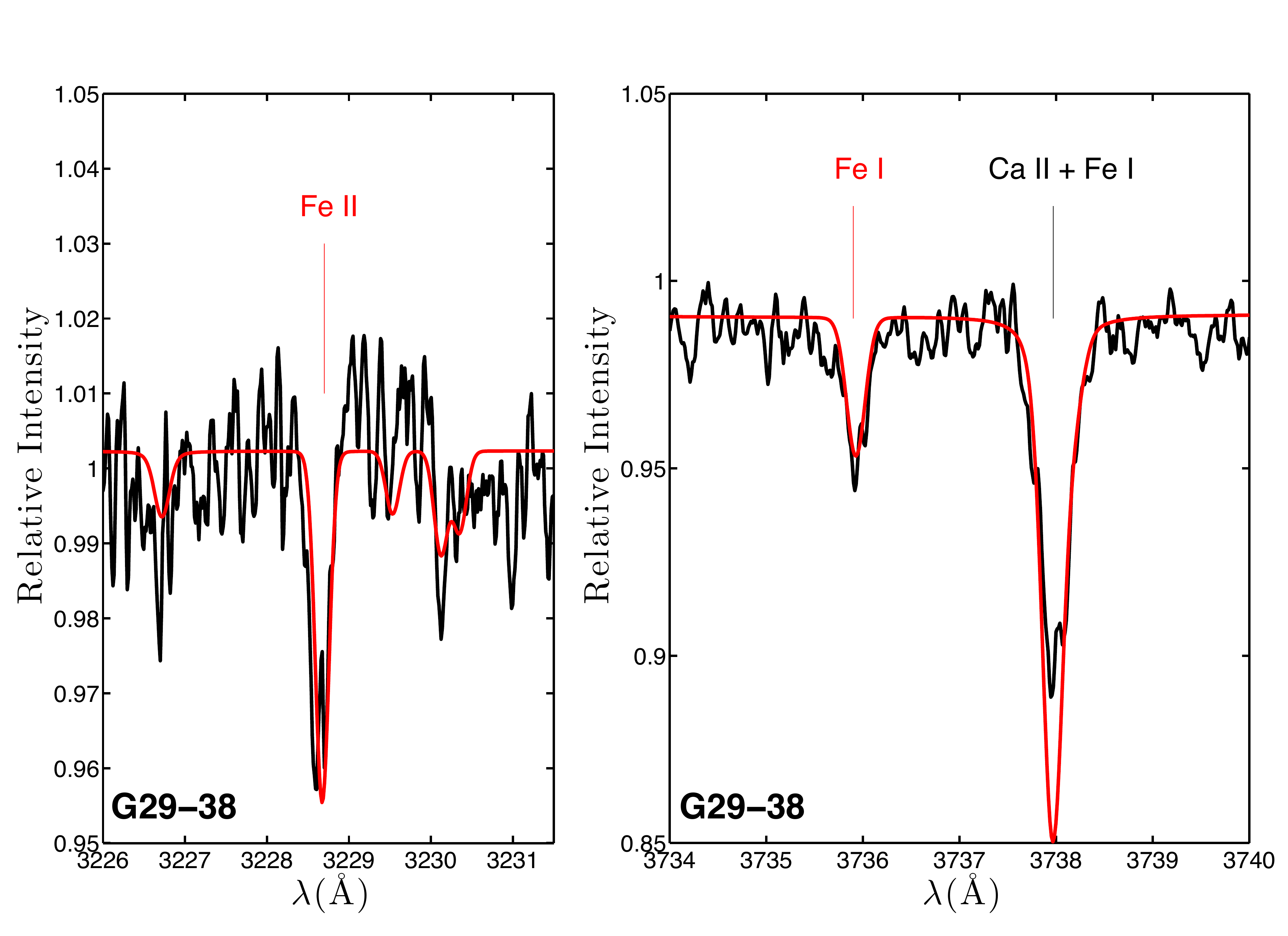}
\caption{Similar to Figure \ref{Fig: C_1} except for iron lines in G29-38. Both spectra were smoothed by a five-point boxcar average. The absorption feature around 3738 {\AA} is a blend of Ca II and Fe I lines, which is dominated by Ca II.}
\label{Fig: Fe_G29-38}
\end{figure}

Three optical iron lines are detected in G29-38 and two of them are presented in Figure \ref{Fig: Fe_G29-38}. Iron is not detected in GD 133. Since molecular hydrogen does not significantly interfere near the Fe II 1361.4 {\AA} region in GD 133, we use the absence of this strong Fe line to derive an upper limit to the abundance. 

\clearpage

\subsection{Additional Upper Limits}

\begin{figure}[hp]
\plotone{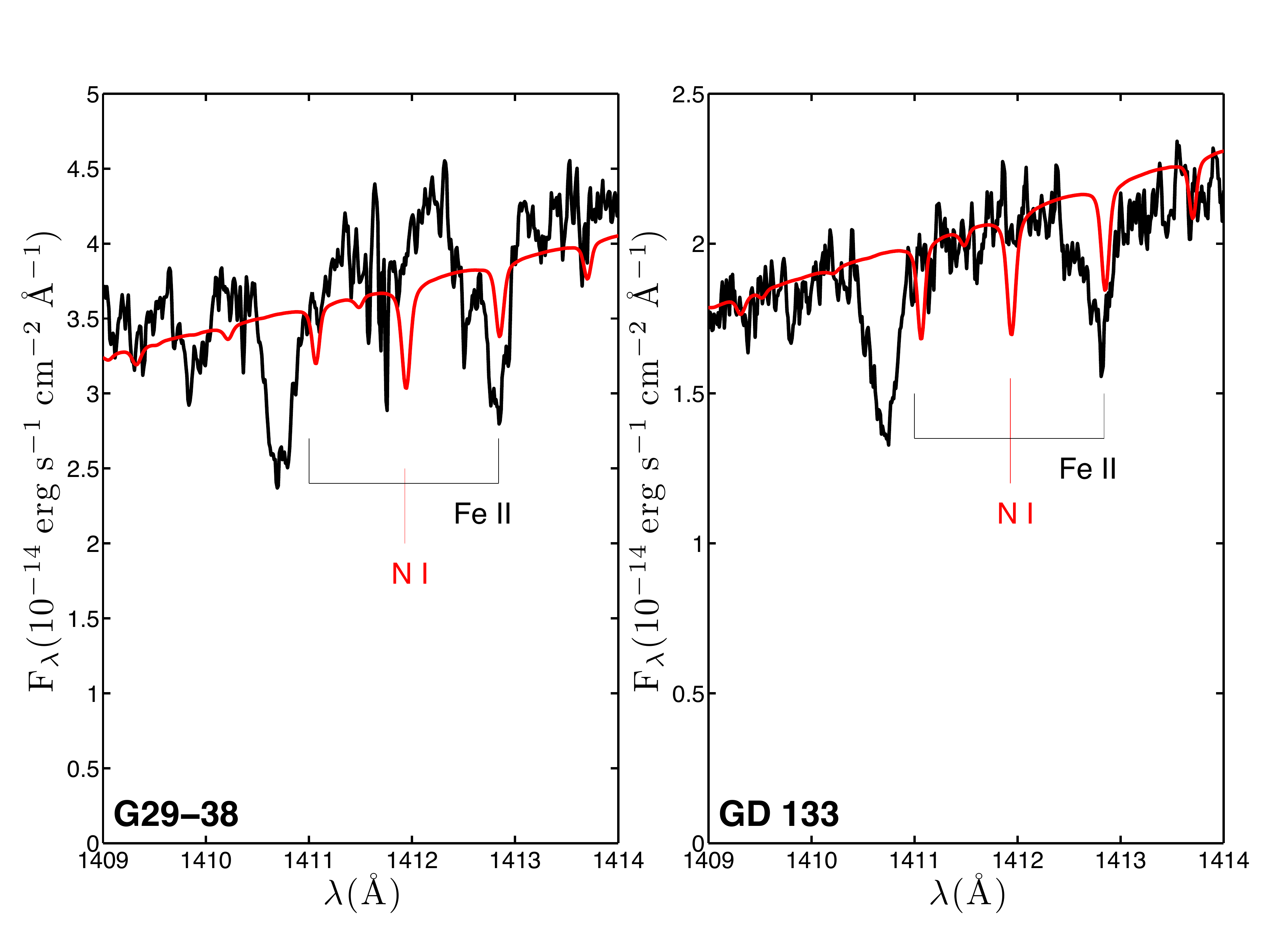}
\caption{Similar to Figure \ref{Fig: C_1} except for N I 1411.9 {\AA}, which is used to derive nitrogen upper limits. Only a model without molecular hydrogen is shown and used for the analysis. All the unlabeled features in the data are from molecular hydrogen.}
\label{Fig: N}
\end{figure}

\begin{figure}[hp]
\plotone{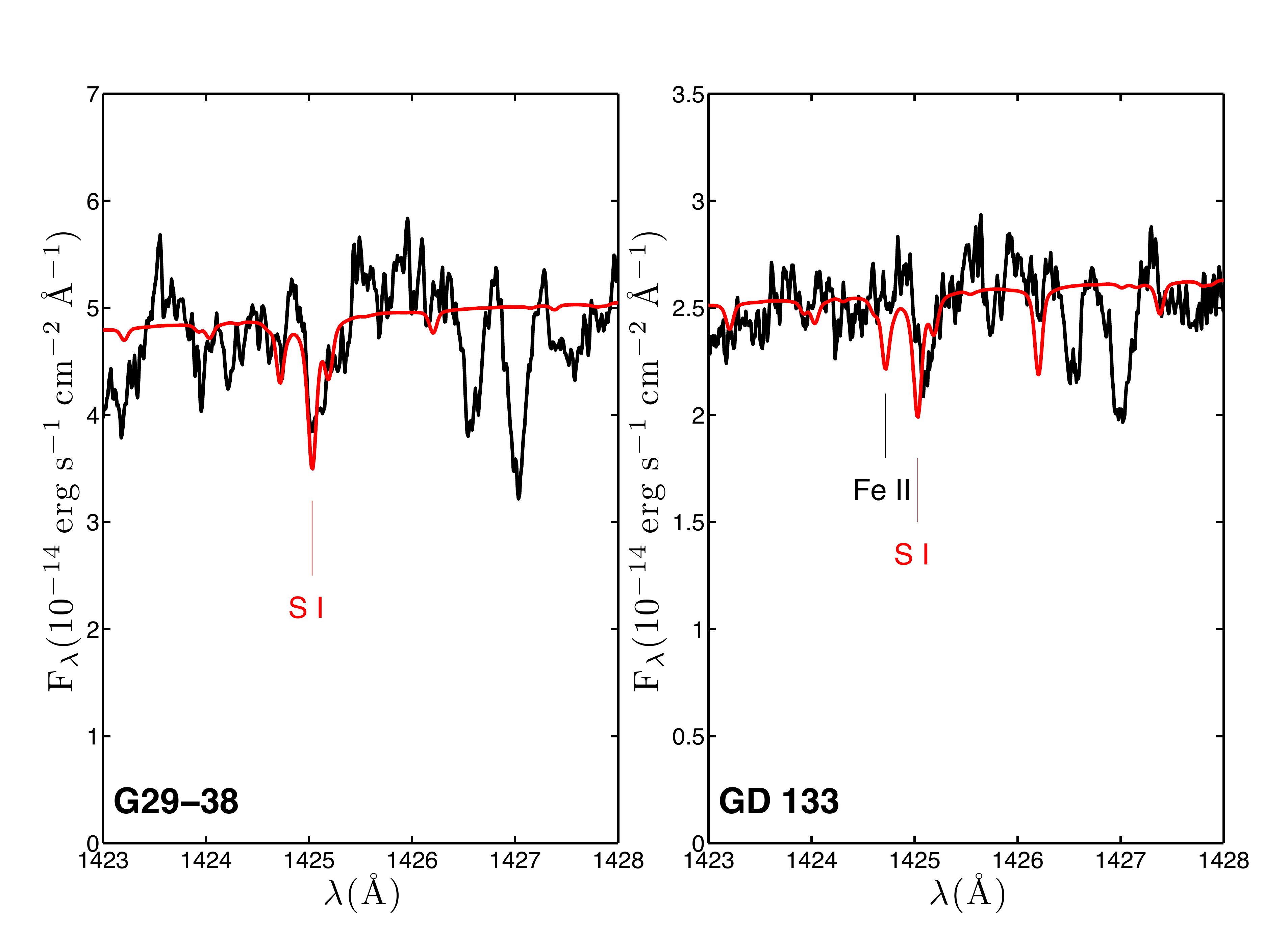}
\caption{Similar to Figure \ref{Fig: C_1} except for S I 1425.0 {\AA}, which is used to derive sulfur upper limits.}
\label{Fig: S}
\end{figure}

{\it Nitrogen, Sulfur, Nickel}: The nitrogen triplet around 1200 {\AA} is in the wing of Lyman $\alpha$ and the S/N of the spectrum is very low, making N I 1411.9 {\AA} the best nitrogen line in our data set, as shown in Figure \ref{Fig: N}. To derive the sulfur upper limit, S I 1425.03 {\AA} was used and the fit is presented in Figure \ref{Fig: S}. Due to the presence of adjacent molecular hydrogen lines, the sulfur upper limit is less constraining in G29-38 than in GD 133. We also used Ni II 1335.2 {\AA} to determine the upper limit as shown in Figure \ref{Fig: C_1}.

{\it Sodium, Aluminum, Manganese}: Optical spectral lines were used to determine the upper limits for these elements. The details are listed in Table \ref{Tab: Lines} and their spectra are not presented.

\section{Discussion}

We have determined the abundances of 8 elements in G29-38, including C, O, Mg, Si, Ca, Ti, Cr, Fe and placed stringent upper limits on S and Ni. There are only three hydrogen dominated white dwarfs with more than 8 elements determined, i.e. 11 for WD 1929+012, 10 for WD 0843+516 \citep{Gaensicke2012} and 9 for NLTT 43806 \citep{Zuckerman2011}. In GD 133, we have detected O, Si and Ca, marginally detected Mg, and placed a meaningful upper limit on the carbon abundance. 

Both G29-38 and GD 133 show excess infrared radiation coming from an orbiting dust disk \citep{ZuckermanBecklin1987, Reach2005b, Reach2009, Jura2007b}. The source of accretion in both G29-38 and GD 133 is likely to come from one large parent body rather than a blend of several small ones, mainly because collisions among different objects will evaporate all the dust particles and destroy the dust disk \citep{Jura2008}. Because the settling times of heavy elements in these white dwarfs are less than a year \citep{Koester2009a}, much shorter than the disk lifetime of $\sim$ 10$^5$ yr \citep{Farihi2012b}, the accretion is likely to be in a steady state, wherein the rate of material falling on the white dwarf atmosphere is balanced by the settling out of the convection zone \citep{Koester2009a}. We can infer the composition of the accreted extrasolar planetesimals based on the accretion flux, which is dependent on the settling times of each element.

\subsection{Calculation of Settling Times}

Very recently, \citet{Deal2013} have drawn attention to the fingering or thermohaline instability; they find that this effect is important in hydrogen dominated white dwarfs and can change the derived accretion rates by orders of magnitude. Unfortunately they do not publish enough details for us to draw conclusions about the relevance of the effect for the white dwarfs studied here. Some problems we see in \citet{Deal2013} are:

\begin{itemize}
\item \citet{Deal2013} use a prescription for the thermohaline diffusion coefficient from \citet{VauclairTheado2012}, which is
claimed to be physically more realistic than previous methods. However, their formula leads to an infinite value for the coefficient at the bottom of the convection zone, which is unrealistic.

\item In our model for G29-38 the bottom of the convection zone is at log $(M_{cvz}/M_{star})$ = -13.9 (see Table \ref{Tab: Parameters}) and the layers above this will be homogeneously mixed in a matter of seconds. In \citet{Deal2013}, the model 3 presented in both Figure 1 and 2 has no convection at all. The model 2, with parameters closest to G29-38 has a convection zone a factor of 100 smaller than G29-38. None of these models can represent our current objects.

\item The H/He interface at log $(M_{cvz}/M_{star})$ = -5 is not a sharp boundary, but a transition zone. The helium will mix upward into the hydrogen and a significant helium fraction will still be present several pressure scale heights above this layer. This will lead to an increasing molecular weight, an effect apparently not considered in \citet{Deal2013}.

\item Their main argument is the fact that in previous determinations the accretion rates for helium white dwarfs seemed to be systematically higher than in hydrogen white dwarfs. In our opinion, a much more likely reason is the uncertainty in the depth of the convection zones in helium white dwarfs, which is described in corrected diffusion times\footnote{see www.astrophysik.uni-kiel.de/$\sim$koester} as well as possibly non-steady state accretion (see section 5.3 for more discussion).
\end{itemize}

For the time being, we do not believe that the \citet{Deal2013} calculations are applicable to G29-38 or GD 133; but we will reconsider our conclusions, once more details become available. 

\subsection{G29-38}

\begin{figure}[hp]
\plotone{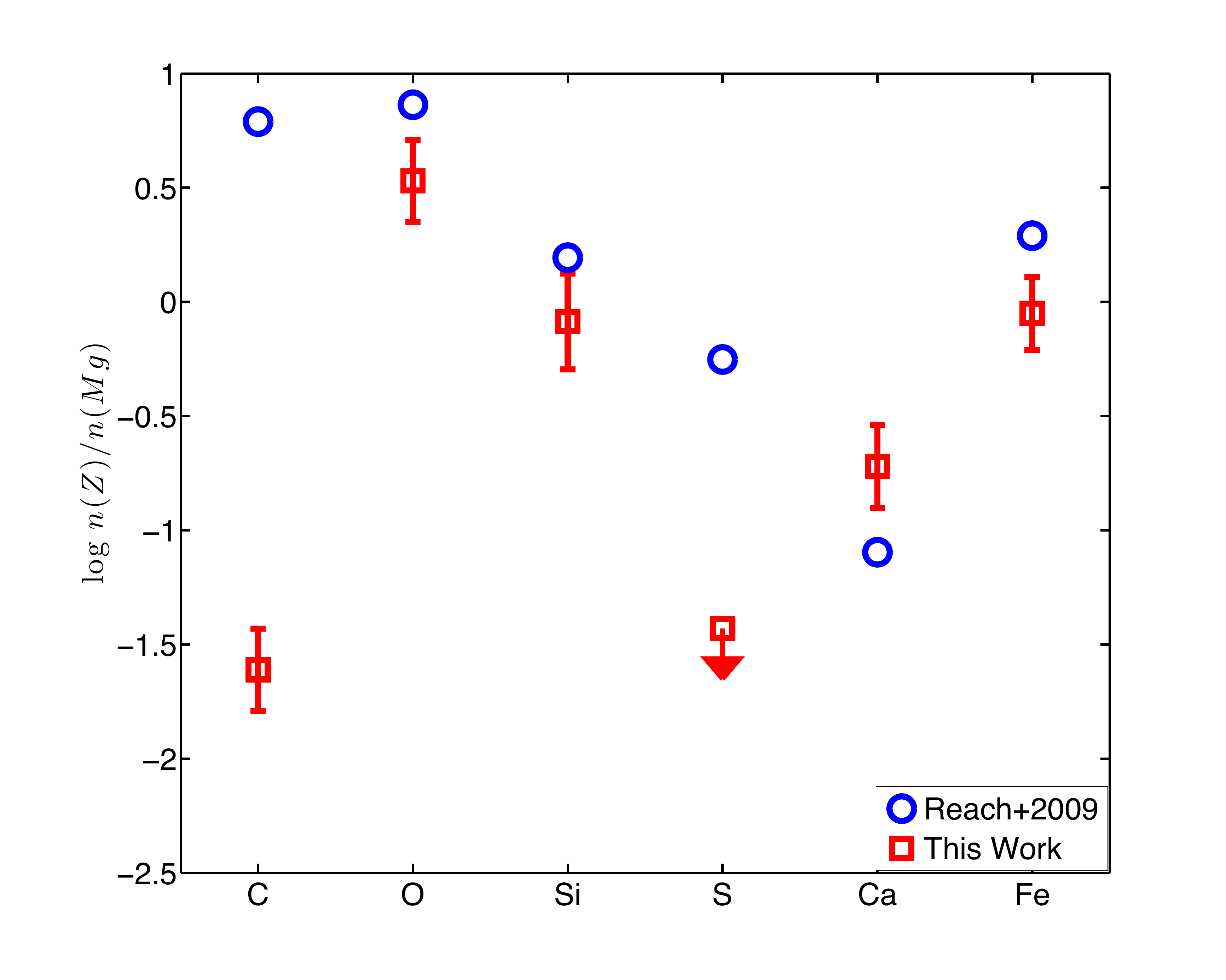}
\caption{A comparison between the composition of the extrasolar planetesimal accreted onto G29-38 derived from atmospheric analysis (from Table \ref{Tab: Abundances}, this work) and fitting the infrared spectrum of the dust disk \citep{Reach2005b, Reach2009}. The atoms are arranged with increasing atomic weight. The ordinate represents the logarithmic value of the number ratios between an element and magnesium, one of the dominant elements. One sigma error bars are plotted. The overall agreement is respectable except for the carbon and sulfur abundances; see section 5.2 for details.}
\label{Fig: comp}
\end{figure}

The abundances in the parent body that accreted onto G29-38 are calculated by correcting for the settling effect of each element in the atmosphere, as shown in Table \ref{Tab: Abundances}. In addition, the abundances can also be derived by fitting the infrared spectrum of the dust disk, which is composed of pulverized extrasolar planetesimals prior to accretion onto the white dwarf. G29-38 has the brightest known dust disk due to its proximity to the Earth and \citet{Reach2005b, Reach2009} found that the dominant minerals are amorphous carbon (C), amorphous and crystalline silicates (MgFeSiO$_4$, Fe$_2$SiO$_4$, Fe$_2$Si$_2$O$_6$, CaMgSi$_2$O$_6$, Mg$_2$Si$_2$O$_6$), water ice (H$_2$O) and metal sulfides (Mg$_{0.1}$Fe$_{0.9}$S). A comparison is shown in Figure \ref{Fig: comp} where we see that the derived number ratios of n(O)/n(Mg), n(Si)/n(Mg), n(Ca)/n(Mg) and n(Fe)/n(Mg) are in rough agreement in the atmosphere and in the surrounding dust disk. The biggest discrepancy is the abundances of carbon and sulfur. 

While \citet{Reach2005b, Reach2009} simulated the disk as an optically thin dust torus, we argue that it is more likely to be mostly opaque as described in \citet{Jura2003, Jura2009a}, for the following reasons. (i) In the optically thin disk model around G29-38, \citet{Reach2009} derived a total disk mass of 2 $\times$ 10$^{19}$ g, which is three orders of magnitude smaller than the lower limit of the mass of heavy elements in the atmosphere of dusty helium dominated white dwarf [see, for example \citet{Jura2012}]. Assuming there is no difference between dusty hydrogen and helium white dwarfs, the disk must be much more massive and therefore optically thick. (ii) For dusty white dwarfs hotter than 20,000 K, the entire optically thin disk would be located outside of the tidal radius \citep{XuJura2012}. There is no viable mechanism to explain the presence of so much hot dust outside the tidal radius of the white dwarf. In the opaque disk model, the radiation from the disk in the main contributor to the continuum, rather than from featureless emissions of minerals. This explains the discrepancy in carbon abundance because the emissivity spectrum of amorphous carbon is featureless \citep{Reach2009}. To account for the strong 10 $\mu$m silicate emission feature, \citet{Jura2009a} proposed the presence of an outer optically thin region or emission from a hot region on top of the opaque disk. Thus, the optically thin model from \citet{Reach2009} for the 10 $\mu$m emission features should work and most of their derived abundances agree with the values in the photosphere. However, the inclusion of niningerite (Mg$_{0.1}$Fe$_{0.9}$S) as the metal sulfide led \citet{Reach2009} to derive a high sulfur abundance; this is inconclusive due to the low S/N of the infrared spectrum (see Figure 5 in \citet{Reach2009}). This is the first direct comparison of elemental abundances derived from fitting the infrared spectrum of a dust disk with the spectroscopic analysis of a white dwarf atmosphere; the overall agreement is respectable.

\begin{figure}[hp]
\epsscale{1.1}
\plotone{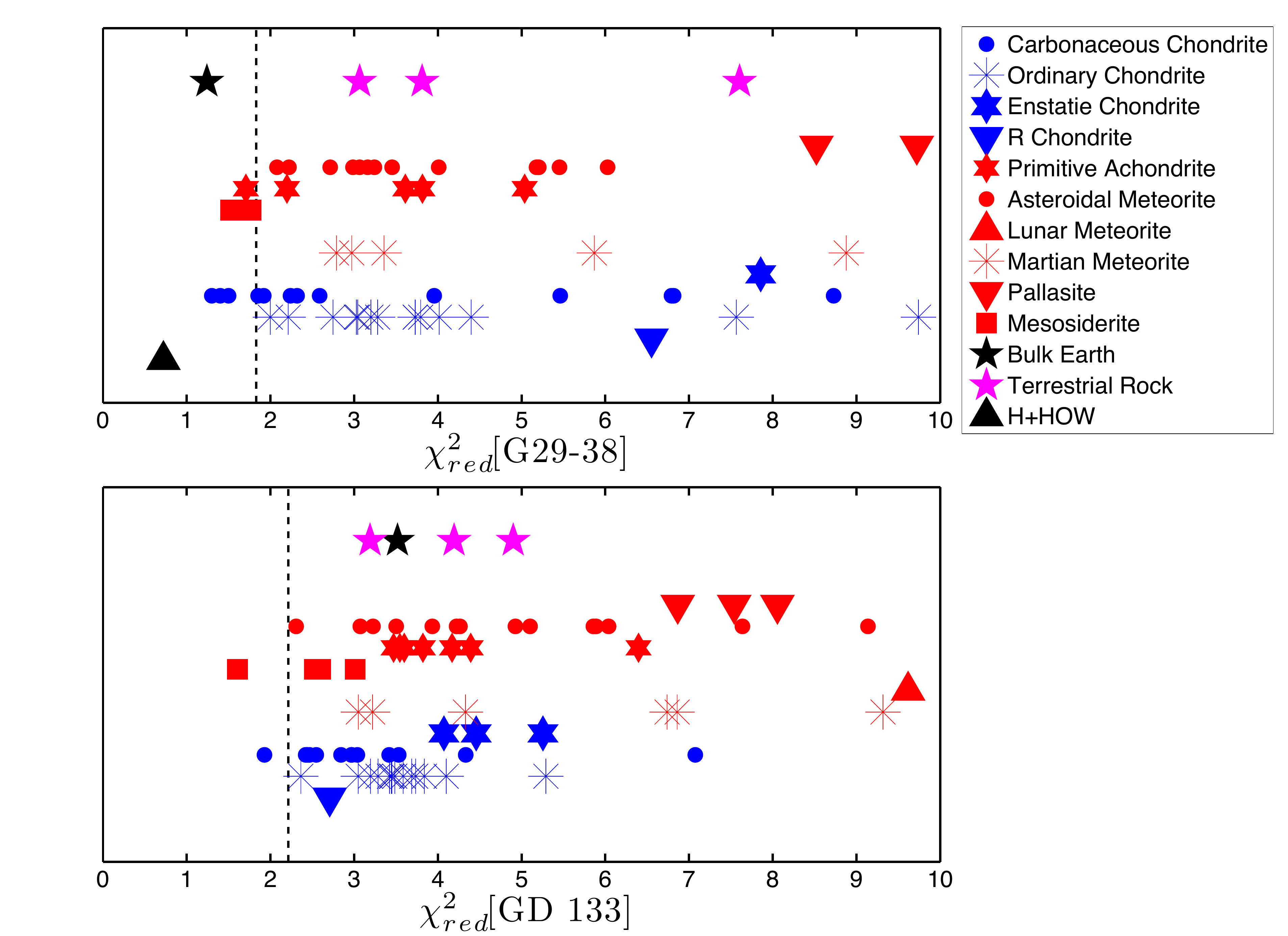}
\caption{ Computed reduced chi-squared values between the observed composition and meteorites. Different meteorite groups are offset in position in the ordinate for clarification. The upper panel is for G29-38 and lower panel for GD 133. For G29-38, we compare the mass fraction of 10 elements (C, O, Mg, Si, S, Ca, Ti, Cr, Fe and Ni) relative to the summed mass of O, Mg, Si and Fe; for GD 133, 5 elements (C, O, Mg, Si and Ca) are considered relative to the summed mass of O, Mg and Si because only an upper limit is obtained for Fe. The dashed lines represent 95\% confidence level. The black triangle represents one of the blends that can best match the abundances observed in G29-38; it consists of 60\% H chondrite and 40\% of howardite. The meteorite database is described in \citet{Xu2013a} with most data from \citet{Nittler2004} and some Martian meteorites data from \citet{McSween1985}. The carbon abundance in CR chondrites and lodranite is from \citet{Alexander2007} and \citet{GradyWright2003}, respectively.  The bulk composition of Earth is from \citet{Allegre2001}. Terrestrial rocks include data from continental crust, upper mantle and lower mantle \citep{Anderson2007}.}
\label{Fig: Chi}
\end{figure}

To find the best solar system analog to the composition of the parent body accreted onto G29-38, we follow \citet{Xu2013a} and perform a reduced chi-squared analysis with different types of meteorites. We consider 10 elements in total, including C, O, Mg, Si, S, Ca, Ti, Cr, Fe and Ni. We assigned the uncertainty in number ratios for S and Ni to be 0.17 dex, the biggest uncertainty from elements with a detection in G29-38. Therefore, these two elements contribute to the reduced chi-squared value but with relatively low weight. As shown in Figure \ref{Fig: Chi}, several single meteorites fall within the 95\% confidence level, including bulk Earth, CR chondrites, primitive achondrites and mesosiderites, a type of stoney-iron meteorites. Mesosiderites are a less promising candidate because the mass fraction of one of the major elements, Mg is 0.3 dex less in mesosiderite than in G29-38. 

\begin{figure}[hp]
\epsscale{1}
\plotone{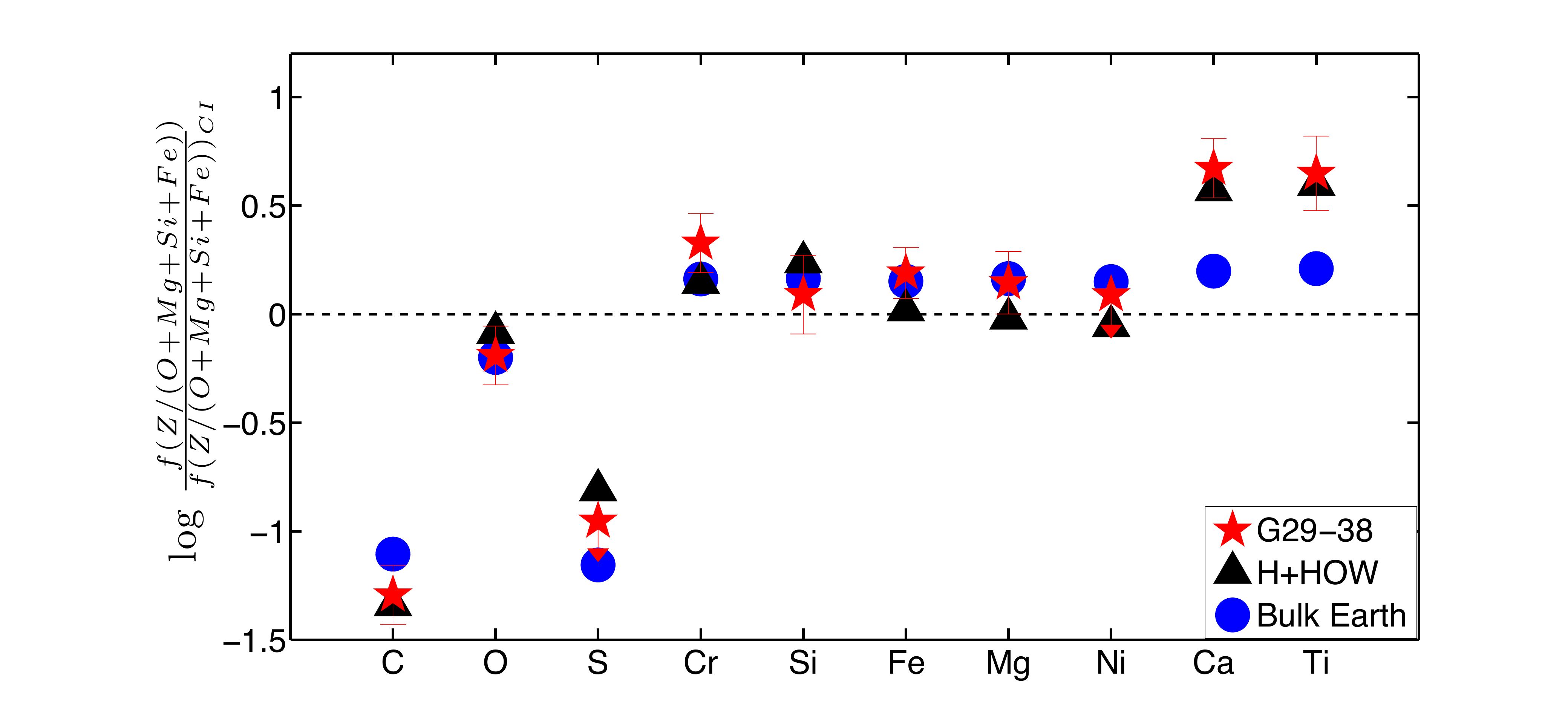}
\caption{For G29-38, mass fraction of an element with respect to the sum of oxygen, magnesium, silicon and iron normalized to that of CI chondrites \citep{WassonKallemeyn1988}. Arrows denotes upper limits. The elements are ordered by increasing condensation temperature.  Also plotted is the best match single solar system object, bulk Earth as well as one of the best match when considering a blend -- 60\% H chondrite and 40\% howadrite. Abundances for G29-38 are taken from Table \ref{Tab: Abundances}.}
\label{Fig: Abundance_G29-38}
\end{figure}

A detailed comparison of all the elements scaled to CI chondrites, the most primitive material in the solar system, is shown in Figure \ref{Fig: Abundance_G29-38}. We see that the mass fraction of volatile elements, such as carbon and sulfur, are depleted by at least a factor of 9 while the refractory elements, including calcium and titanium, are enhanced. The composition of the best-match solar system object -- bulk Earth -- is also plotted for comparison. The largest discrepancies between the compositions of the object accreted onto G29-38 and bulk Earth are Ca and Ti, two refractory elements. 

When considering a blend of two meteorites, several combinations can all provide a good fit to the composition observed in G29-38. One example includes 60\% H chondrite and 40\% howardite, as shown in Figures \ref{Fig: Chi} and \ref{Fig: Abundance_G29-38}. In general, some refractory-enhanced achondrites, such as howardites or mesosiderites, are required to reproduce the refractory abundance and chondritic material is needed to adjust the overall abundance pattern. Though the exact mechanism is not known, it is clear that the parent body accreted onto G29-38 has experienced post-nebular processing, such as differentiation and collision, which is found to be common for extrasolar planetesimals \citep{Xu2013a}.

G29-38 has accreted from an extrasolar planetesimal which is enhanced in both calcium and titanium, which is less frequently seen than enhancement of only calcium. In a sample of well studied polluted white dwarfs compiled in \citet{JuraYoung2014}, 6 out of 12 stars have [Ca/Mg] at least a factor of two higher than the value for CI chondrites; in comparison, only 3 out of 9 stars have a factor of two [Ca/Mg] and [Ti/Mg] enhancement, including PG 1225-079, GD 362 and G29-38. These three stars are the best candidates for accretion of a ``normal" object and a refractory-dominated planetesimal \citep{JuraXu2013}. In addition, both G29-38 and GD 362 have well constrained stellar masses from parallax measurements \citep{vanAltena2001, Kilic2008b}. Using the initial mass to final mass relationship \citep{Williams2009}, we derive a progenitor mass of 3.95M$_\odot$ and 2.95M$_{\odot}$ for G29-38 and GD 362, respectively, which are higher than the average white dwarf progenitor mass \citep{Kleinman2013}. Unfortunately, PG 1225-079 is in the stellar parameter region where temperature and surface gravity are coupled when using only the spectroscopic method \citep{Klein2011}.  At least 2 out of the 3 stars that have accreted from refractory-dominated planetesimals have high progenitor mass; this is consistent with the model that refractory-dominated planetesimals are more likely to survive the red giant stage of stars with relatively high main-sequence masses \citep{JuraXu2013}.

As shown in Figure \ref{Fig: Abundance_G29-38}, the oxygen abundance in G29-38 is depleted by a factor of 1.5 compared to that in CI chondrites. The total amount of oxygen is barely enough to combine with all the heavy elements into the oxides; there might have been some metallic iron in the accreted planetesimal. No excess oxygen is left to be in the form of H$_2$O and water is very depleted in the planetesimal accreted onto G29-38.

Though \citet{Reach2009} derived a high water abundance, it is not supported by the NIRSPEC data shown in Figure \ref{Fig: NIRSPEC}. No emission lines from water are detected. We see a gentle slope towards longer wavelength, which is consistent with the dust disk model, peaking longward of 4 $\mu$m. The IRAC 3.6 $\mu$m flux of G29-38 is $\sim$ 10\% higher than predicted by the model and \citet{Farihi2008a} hypothesized contributions from some PAH features. A subsequent study by \citet{Reach2009} found that there is significant fluctuation at 3.6 $\mu$m with an amplitude of $\sim$ 5\%, which could account for this discrepancy.  Our NIRSPEC data also exclude any circumstellar emissions from PAH features as observed in Herbig Ae/Be stars \citep{Meeus2001}.

\subsection{GD 133}

For GD 133, we also performed a reduced chi-squared analysis for 5 heavy elements, including C, O, Mg, Si and Ca with an uncertainty of 0.2 dex for C and Mg. Their mass fractions are calculated with respect to the summed mass of O, Mg and Si, the most abundant three elements in GD 133. As shown in Figure \ref{Fig: Chi}, CR chondrites and mesosiderites match with the abundance pattern observed in GD 133. 

\begin{figure}[hp]
\plotone{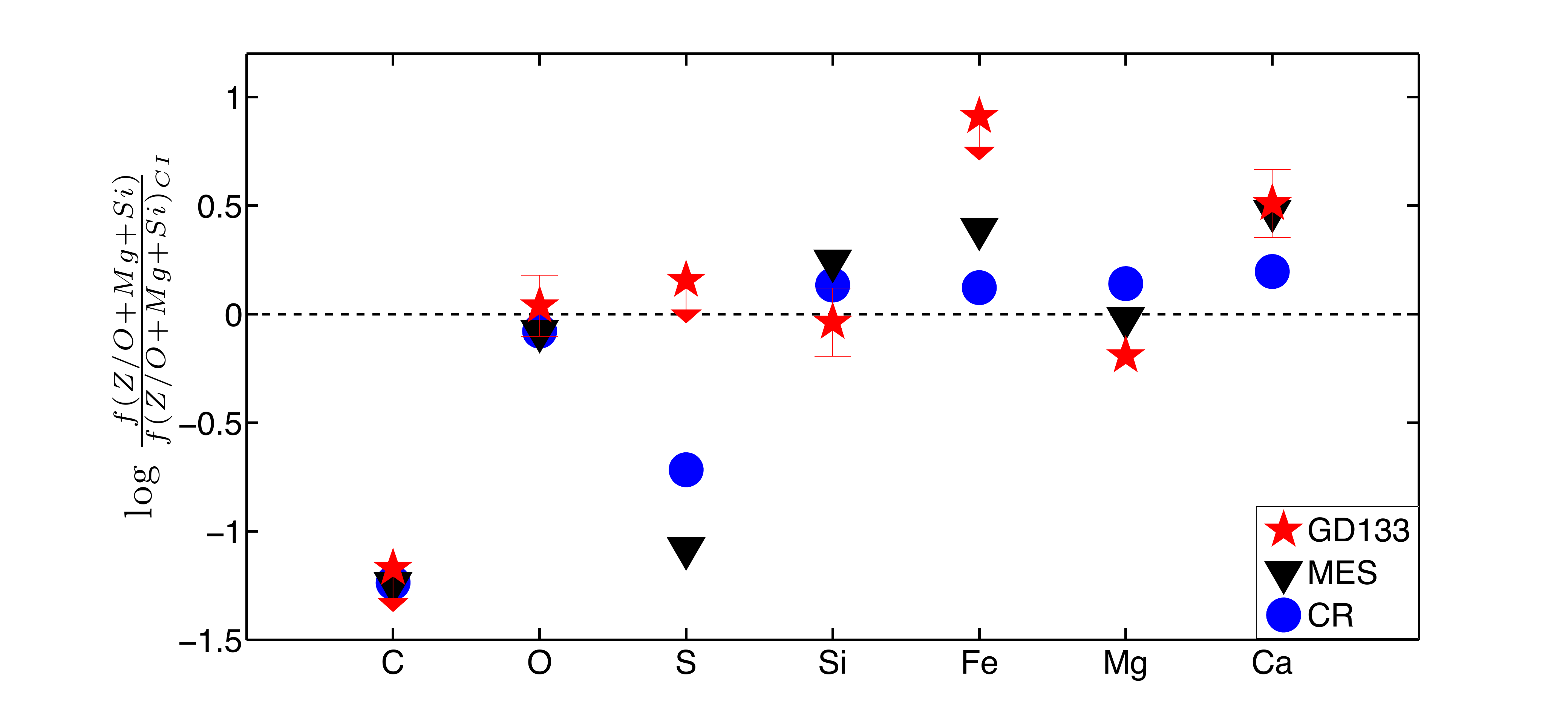}
\caption{Similar to Figure \ref{Fig: Abundance_G29-38} except for GD 133 and the mass fraction of an element normalized to the sum of oxygen, magnesium and silicon. There is no error bar associated with magnesium because it is only marginally detected. The compositions of two best match meteorites from the reduced chi-squared analysis, mesosiderite and CR chondrite, are also plotted for comparison. }
\label{Fig: Abundance_GD133}
\end{figure}

A detailed comparison of the composition of the parent body accreted onto GD 133 and solar system objects is shown in Figure \ref{Fig: Abundance_GD133}. Compared to CI chondrites, carbon is depleted by at least a factor of 20 in GD 133 while calcium is enhanced by a factor of 3. Magnesium is slightly depleted and [Mg/Ca] = 0.54, which is on the high side of all polluted white dwarfs \citep{JuraXu2013}. Mesosiderites give a better fit to the overall abundance pattern with the similar Ca enhancement and Mg depletion. \citet{Xu2013a} also found that mesosiderites is the best match to the abundance pattern in GD 362 due to the enhancements of Ca, Ti and Al. With constraints from only 5 elements in GD 133, it is hard to derive additional information. There exist several strong Fe II and Mg II lines around 2800 {\AA} and they will provide much more information about the nature of the accreted material.

Until now, all dusty white dwarfs are heavily polluted with mass accretion rates of at least 1 $\times$ 10$^8$ g s$^{-1}$ \citep{Farihi2009, Brinkworth2012}. GD 133 has a dust disk, which reprocesses about 0.5\% of the incoming star light \citep{Jura2007b,Farihi2010b}. Assuming a chondritic iron to silicon ratio, the accretion rate of iron is 5.9 $\times$ 10$^6$ g s$^{-1}$ and the total accretion rate is 3.0 $\times$ 10$^7$ g s$^{-1}$, the lowest of all dusty white dwarfs. Even with an iron abundance of [Fe/H] = -5.90 (the upper limit listed in Table \ref{Tab: Abundances}), the total accretion rate can only add up to 7.6 $\times$ 10$^7$ g s$^{-1}$, marginally lower than all other dusty white dwarfs. In a steady state model, Poynting-Robertson drag provides a lower bound of the accretion rate \citep{Rafikov2011a, XuJura2012}:

\begin{equation}
\dot{M}=\frac{16\phi_r}{3}\frac{r_*^3}{r_{in}}\frac{\sigma T_{eff}^4}{c^2}
\end{equation}
$\phi_r$ is an efficiency coefficient and taken as 1. $\sigma$ and c are the Stephan-Boltzman constant and the speed of light, respectively. For GD 133, the stellar temperature T$_{eff}$ is listed in Table \ref{Tab: Parameters} and the stellar radius r$_*$=0.012r$_\odot$. Depending on the disk inclination, the inner radius of the disk can vary \citep{Jura2007b}. To derive a lower limit of the mass accretion rate, we take the largest possible inner radius r$_{in}$=23r$_*$ for a nearly face-on disk, and consequently $\dot{M}$ = 2.5 $\times$ 10$^8$ g s$^{-1}$, about an order of magnitude higher than the most likely inferred value. The exceptionally low accretion rate in GD 133 provides direct evidence that equation (1) does not always hold, possibly because the accretion process onto the star is not always in a steady state. 

There exists additional evidence for time-varying non-steady state accretion. Based on the different accretion rates derived from hydrogen and helium atmosphere white dwarfs, \citet{Farihi2012b} postulated a non-steady state accretion model including a high accretion stage and a low accretion stage. Now with an updated helium model atmosphere and settling times \citep{Xu2013a}, the difference has narrowed but is still present.

\section{Perspective}

\begin{figure}[hp]
\plotone{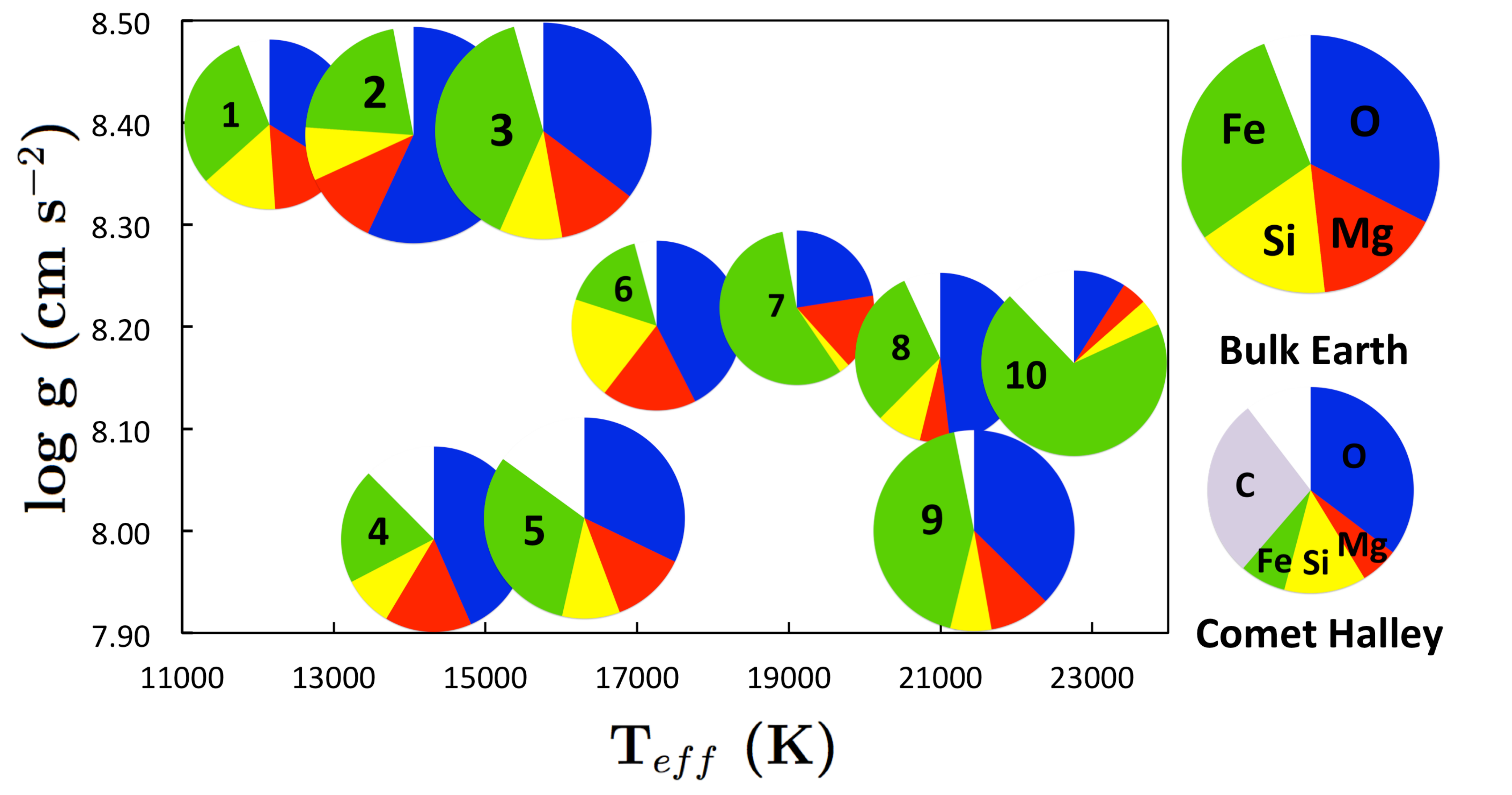}
\caption{A compilation of all polluted white dwarfs with detections of at least O, Mg, Si and Fe; 8 of these stars also have good constraints on the carbon abundance. The abscissa marks white dwarf effective temperatures, corresponding to a cooling age less than 500 Myr; the ordinate is surface gravity, which shows a main sequence mass between 1.8-4.0 M$_{\odot}$ for the current sample. For clarity, some objects are slightly offset in their plotted positions. The abundances have been corrected for the effect of settling and we only show the mass fraction of O, Mg, Si and Fe; the rest are left blank. The size of a pie correlates with the accretion rate (not to scale). We see that O, Mg, Si and Fe are always the dominant elements in a variety of extrasolar planetesimals, resembling bulk Earth. No carbon rich planetesimals similar to comet Halley have been identified so far. The white dwarfs are ordered with increasing stellar temperatures. {\bf References:} Hydrogen dominated white dwarfs: 1: G29-38 (this paper), 7: PG 1015+161, 8: WD 1226+110, 9: WD 1929+012, 10: WD 0843+516 \citep{Gaensicke2012}; helium dominated white dwarfs: 2: WD J0738+1835 \citep{Dufour2012}, 3: HS 2253+8023 \citep{Klein2011}, 4: G241-6, 5: GD 40 \citep{Jura2012}, 6: GD 61 \citep{Farihi2011a, Farihi2013}. All white dwarfs  except \# 3 and 4 have a dust disk. Bulk Earth: \citet{Allegre2001}. Comet Halley: \citet{Jessberger1988}}
\label{Fig: Pie}
\end{figure}

So far, there are 10 white dwarfs that have detections of at least O, Mg, Si and Fe in the atmosphere.  As presented in Figure \ref{Fig: Pie}, they are sampled from a variety of stellar temperatures and surface gravity.  Five of these stars have a hydrogen-dominated atmosphere and five of them have a helium-dominated atmosphere. Eight stars also have a dust disk. Yet, to zeroth order, the resemblance of their elemental compositions relative to bulk Earth is robust.  No carbon rich extrasolar planetesimals, e.g. analogs to comet Halley with 28\% carbon \citep{Jessberger1988} or interplanteary dust particle with $\sim$ 10\% carbon \citep{Thomas1993}, are identified in the current sample. The only white dwarf that has accreted objects with a considerable amount of water is GD 61 (\# 6 in the plot), which contains 26 \% water by mass \citep{Farihi2013}. The general properties of extrasolar planetesimals can be summarized as the following \citep{JuraYoung2014}, (i) Oxygen, iron, silicon and magnesium always dominate and make up more than 85\% of the total mass; (ii) carbon is always depleted relative to the solar abundance; (iii)  viewed as an ensemble, water is less than 1\% of the total accreted mass but exceptions exist. As shown in Figure \ref{Fig: Pie}, there are variations among the abundances of different elements but they are only a factor of 2-3, comparable to the errors. To move beyond a zeroth order result, one must determine abundances of as many trace elements as possible, such as the case of GD 362 \citep{Zuckerman2007, Xu2013a}. Alternatively, one can assess the abundances of a few elements in an ensemble of stars, e.g. \citet{JuraXu2012, JuraXu2013}.

\section{Conclusions}

In this paper, we report optical and ultraviolet spectroscopic studies of two externally-polluted hydrogen dominated white dwarfs, G29-38 and GD 133. For G29-38, with the exception of carbon and sulfur, the derived elemental abundances agree reasonably well with the values obtained from fitting the infrared spectrum of the dust disk. Both stars have accreted objects that show a pattern of volatile depletion and refractory enhancement. The parent body accreted onto G29-38 has experienced post-nebular processing and can be best explained by a blend of chrondritic material and a refractory-enhanced object. The total mass accretion rate in GD 133 is significantly lower than all other dusty white dwarfs, suggesting non-steady state accretion. In a sample of ten stars, we find that the elemental compositions of extrasolar planetesimals are similar to bulk Earth regardless of their evolutionary history.

We thank G. Mace for helping with the NIRSPEC observing run and useful discussions about data reduction procedures, C. Melis for helping with HIRES observing runs in 2008, and B. Holden for useful email exchanges regarding the MAKEE software. Support for program \#12290 was provided by NASA through a grant from the Space Telescope Science Institute, which is operated by the Association of Universities for Research in Astronomy, Inc., under NASA contract NAS 5-26555. The authors wish to recognize and acknowledge the very significant cultural role and reverence that the summit of Mauna Kea has always had within the indigenous Hawaiian community.  We are most fortunate to have the opportunity to conduct observations from this mountain. This work has been partly supported by NSF \& NASA grants to UCLA to study polluted white dwarfs.

\bibliographystyle{apj}
\bibliography{apj-jour,WD.bib}

\end{CJK}
\end{document}